\documentclass[aps,12pt,final,notitlepage,oneside,onecolumn,%
nobibnotes,nofootinbib,superscriptaddress,noshowpacs,centertags]{revtex4}
\usepackage{epsfig, graphicx, euscript}
\usepackage{afterpage}

\def\l{{\ell}}
\def\lm{{\l m}}

\def\alm{a_{\lm}}
\def\ylm{Y_{\lm}}

\def\healpix{H{\sc ealpix }}
\def\glesp{G{\sc lesp }}

\def\etal{et al.}

\def\lmax{{\l_{\rm max}}}

\def\ijmpd{\it Int. J. Mod. Phys. D}
\def\aaa{\it A \& A}

\newcommand{\nbi}{{Niels Bohr Institute, Blegdamsvej 17,
DK-2100 Copenhagen, Denmark}}
\newcommand{\sao}{{Special Astrophysical Observatory, Nizhnij Arkhyz,
Karachaj-Cherkesia, 369167, Russia}}

\newcommand{\imperial}{{Imperial College, London, United Kingdom}}
\newcommand{\cardiff}{{School of Physics \& Astronomy, Cardiff University,
    5 The Parade, Cardiff, CF24 3AA, Wales, United Kingdom}}

\begin{document}

\setcounter{page}{101}

\title{Understanding the WMAP Cold Spot mystery
\footnote{Astrophysical Bulletin {\bf 65}, Issue 2, pp.101--120}.}

\author{\firstname{P.~D.}~\surname{Naselsky}}
\email{naselsky@nbi.dk}
\author{\firstname{P.~R.}~\surname{Christensen}}
\email{perrex@nbi.dk}
\affiliation{\nbi}
\author{\firstname{P.}~\surname{Coles}}
\email{peter.coles@astro.cf.ac.uk}
\affiliation{\cardiff}
\author{\firstname{O.~V.}~\surname{Verkhodanov}}
\email{vo@sao.ru}
\affiliation{\sao}
\author{\firstname{D.~I.}~\surname{Novikov}}
\email{d.novikov@imperial.ac.uk}
\affiliation{\imperial}
\author{\firstname{Jai.}~\surname{Kim}}
\email{jkim@nbi.dk}
\affiliation{\nbi}

\begin{abstract}
The first and third year data releases from the Wilkinson Microwave
Anisotropy Probe (WMAP) provide evidence of an anomalous Cold Spot
(CS) at galactic latitude $b=-57^\circ$ and longitude $l=209^\circ$.
We have examined the properties of the CS in some detail in order to
assess its cosmological significance. We have performed a cluster
analysis of the local extrema in the CMB signal to show that the CS
is actually associated with a large group of extrema rather than
just one.
We have also
checked the idea that the CMB signal has a non-Gaussian tail.
For each ring we apply a linear filter with
characteristic scale $R$, dividing the CMB signal in two parts: the
filtered part, with characteristic scale above that of the filter
$R$, and the difference between the initial and filtered signal.
Using the filter scale as a variable, we can maximize the skewness
and kurtosis of the smoothed signal and minimize these statistics
for the difference between initial and filtered signal. We find
that, unlike its Northern counterpart, the Southern Galactic
hemisphere of the CMB map is characterized by significant departure
from Gaussianity of which the CS is not the only manifestation: we
have located a ring, on which there are ``cold'' as ``hot'' spots
with almost the same properties as the CS. Exploiting the similarity
of the WCM and the ILC maps, and using the latter as a guide map, we
have discovered that the shape of the CS is formed primarily by the
components of the CMB signal represented by multipoles between
$10\le \ell\le 20$, with a corresponding angular scale about
$5-10^\circ$. This signal leads to modulation of the whole CMB sky,
clearly seen at $|b|>30^\circ$ in both the ILC and WCM maps, rather
than a single localized feature. After subtraction of this
modulation, the remaining part of the CMB signal appears to be
consistent with statistical homogeneity and Gaussianity. We
therefore infer that the mystery of the WMAP CS reflects directly
the peculiarities of the low-multipole tail of the CMB signal,
rather than a single local (isolated) defect or manifestation of a
globally anisotropic cosmology.
\end{abstract}

\maketitle

\section{Introduction}

An extremely Cold Spot (CS), apparently inconsistent with the
assumption of statistically homogeneous Gaussian fluctuations, was
detected in a wavelet analysis
\cite{santanderng,cruz2005,cjt}   
of the
first-year data release from the Wilkinson Microwave Anisotropy
Probe (WMAP). More recently, the existence of this spot has been
confirmed
\cite{cruz2007a,cruz2008}       
the WMAP third year data release
\cite{wmap3ytem,wmap3ycos}.       
The WMAP CS is centered
at the position $b=-57^\circ,l=209^\circ$ in Galactic Coordinates and
has a characteristic scale about $10^\circ$.
As was pointed out
Cruz et al.
\cite{cruz2006},             
the frequency dependence of the signal in the spot area is extremely
flat. This fact has been used by the authors mentioned above to
argue that the WMAP CS  belongs to the CMB signal, rather than any
form of  foreground emission. Cruz et al. pointed out that
the reason the CS was not been detected in real space before the
wavelet analysis was that it was hidden amongst structures at
different scales.

As an apparent example of non-Gaussian behavior in the WMAP CMB
signal, the CS has attracted very serious attention from the
theoretical point of view.
Tomita
\cite{tomita}         
suggested that the CS can be related to second-order gravitational
effects.
Inoue and Silk
\cite{insilk}    
proposed a model involving local
compensated voids. The origin of the CS in connection to the
brightness and number of counts of the NVSS
\cite{nvss}
sources (smoothed on the
scale of a  few degrees) was recently discussed by Rudnick, Brown and
Williams
\cite{insilk}.    
They have detected a 20-45$\%$  dip in the smoothed
NVSS source counts which can be interpreted, they argue, as a
manifestation of the integrated Sachs-Wolfe effect, seen for a
single region of the CMB sky.
Jaffe et al.        
\cite{jaffebianchi,jaffebian2},
Cayon et al.        
\cite{no_crit}
and McEwen et al.   
\cite{mcewen,mcewenwmap3}
have investigated the Bianchi VII$_h$ anisotropic cosmological model
as a possible explanation of the CS and other features of the WMAP
low multipoles. Recently,
Cruz et al.         
\cite{cruz2007b,cruz2008}
have pointed out that  the CS could be produced by a cosmic texture,
assuming that the CMB signal is a combination of the Gaussian and
non-Gaussian parts. The present status
of the problem of existence of the CS remains uncertain, despite the
presence of a vast collection of theoretical suggestions.
Nevertheless, if we believe that one particular part of the WMAP CMB
signal contains non-Gaussian features, it would be necessary to seek
corroborating evidence of non-Gaussianity elsewhere in order to
understand their properties more fully. In this Paper we therefore
present a detailed investigation of the properties of the CS,
focusing attention on the following topics.

First, in Section 2, we show how the CS can be easily detected in
the pixels domain not only in the derived CMB signal, but even in
the WMAP maps for K-W bands before separation of the signal into CMB
and foreground components.

Second, we will demonstrate that the CS belongs to a cluster of
local minima, the spatial distribution of which is modulated by the
large-angle modes of the CMB signal outside the Galactic plane. For
that we use the Internal Linear Combination (ILC) III map and the
co-added WCM map with $N_{side}=512$ in the HEALPix format,
converted to GLESP format
\cite{glesp},      
where each
iso-latitude ring has the same number of pixels $N_{\phi}=2048$ in
azimuthal direction  $\phi$ in polar coordinates. After that we
perform a cluster analysis
\cite{novjorg}      
of the
positive and negative peaks for  selected rings in the area outside
the Kp0 mask, mostly concentrating our attention on the ring
crossing the CS at its extrema $b=-57^\circ$ and $-180^\circ\le
\phi\le 180^\circ$. Taking into consideration the signal for each
ring with the latitude $b$, we can investigate the morphology of the
CMB signal at each latitude for the whole range of $\phi$. This
approach allows to connect the morphology of the CS to the signal
outside the CS for the same latitude $b=-57^\circ$. We will show
that the cluster contained the CS, is not a unique feature of the
$b=-57^\circ$ iso-latitude ring. For example, close to the CS there
are two significant clusters of maxima, but these peaks have lower
amplitude that the CS.

 Next, since the origin of large clusters of extrema
 is related to the angular modulation of the signal on large scales
\cite{novjorg},        
we  split the CMB signal into two
parts. To do that we use the skewness and the kurtosis of the signal
for selected rings, including  the $b=-57^\circ$ ring. Then by using
a simple linear smoothing  filter with characteristic scale $R$ we
separate  the signal into a smoothed component and to a difference
between initial signal and the smoothed component.
 For the smoothed signal  we define the skewness $S(R)$ and the kurtosis
$K(R)$ as a functions of $R$ and find the scale of filtering which
maximize both these characteristics
$S(R_{opt}),K(R_{opt})\rightarrow \max$. By using this scale
$R_{max}$ we separate the initial CMB signal in two parts, one of
them (the smoothed one) contain a maximally non-Gaussian signal, and
the another one (initial signal minus the smoothed one) the
maximally Gaussian signal. The non-Gaussian part is mainly formed by
the signal localized at the range of multipoles $2\le \ell\le 20$
and the other one belongs to the $\ell>20$ multipoles.
 Our analysis clearly demonstrates that the pronounced non-Gaussianity of
the CS reflects directly the existence of a large-scale angular
modulation of the CMB signal with $10\le \ell\le 20$.

Finally, using cluster analysis in combination with skewness and
kurtosis statistics, we are able to detect a few additional cold and
hot spots on the same $b=-57^\circ$ ring as the famous one. To show
that the effect on clustering of the peak by low multipoles  of the
CMB is very common, we took into consideration the north Galactic
hemisphere with deficit of the power and have found a few cold and
hot spots. The idea of implementing of cluster analysis in
combination with skewness and kurtosis statistics  was stimulated
in
\cite{santanderng,cruz2005,cruz2006,cruz2007a,cjt},
and especially
in
\cite{eriksenasym}        
and
\cite{larson}.            

In our analysis we use both the WCM map and the  ILC third year map,
which are very similar outside the $b=\pm 25^\circ$ cut of the
Galactic plane. For our analysis we use the high resolution ILC III
map as a guide map to mark possible zones of the CMB sky in which
the enhanced clustering of the peaks is expected to be considerable.

\section{``Naive'' detection of the WMAP Cold Spot}

\begin{figure}[!th]
\setcaptionmargin{5mm}
\onelinecaptionstrue
\centering \vspace{0.01cm}\hspace{-0.1cm}\epsfxsize=16cm
\epsfxsize=0.6\columnwidth \epsfbox{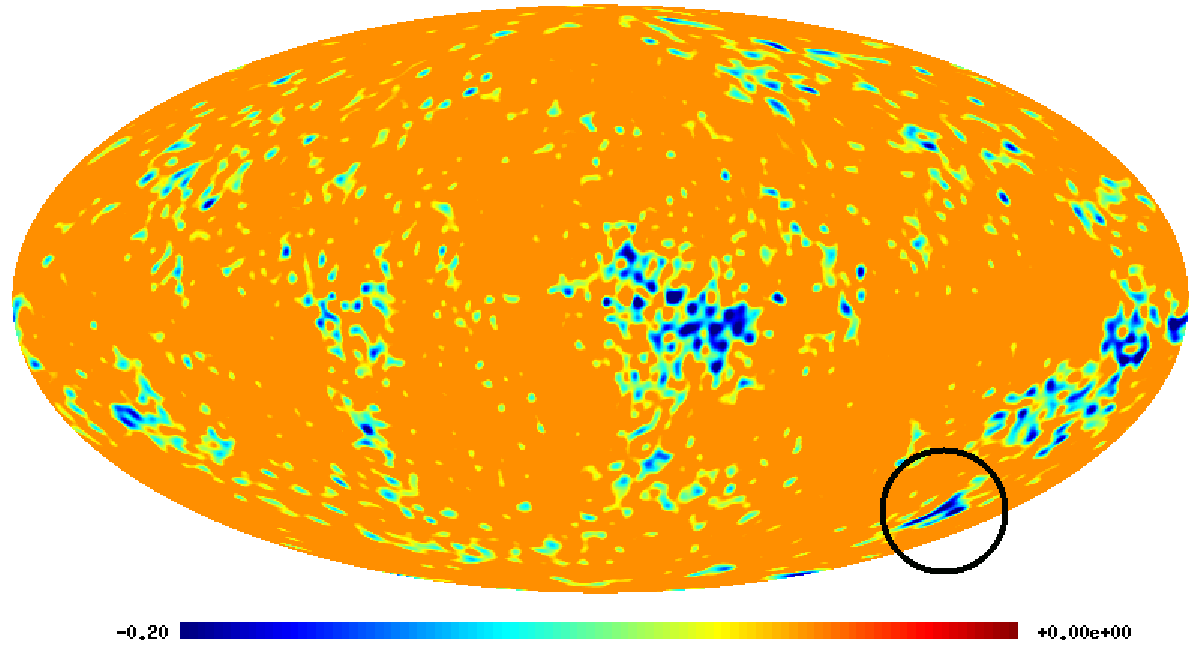} \centering
\vspace{0.01cm}\hspace{-0.1cm}\epsfxsize=16cm
\epsfxsize=0.6\columnwidth \epsfbox{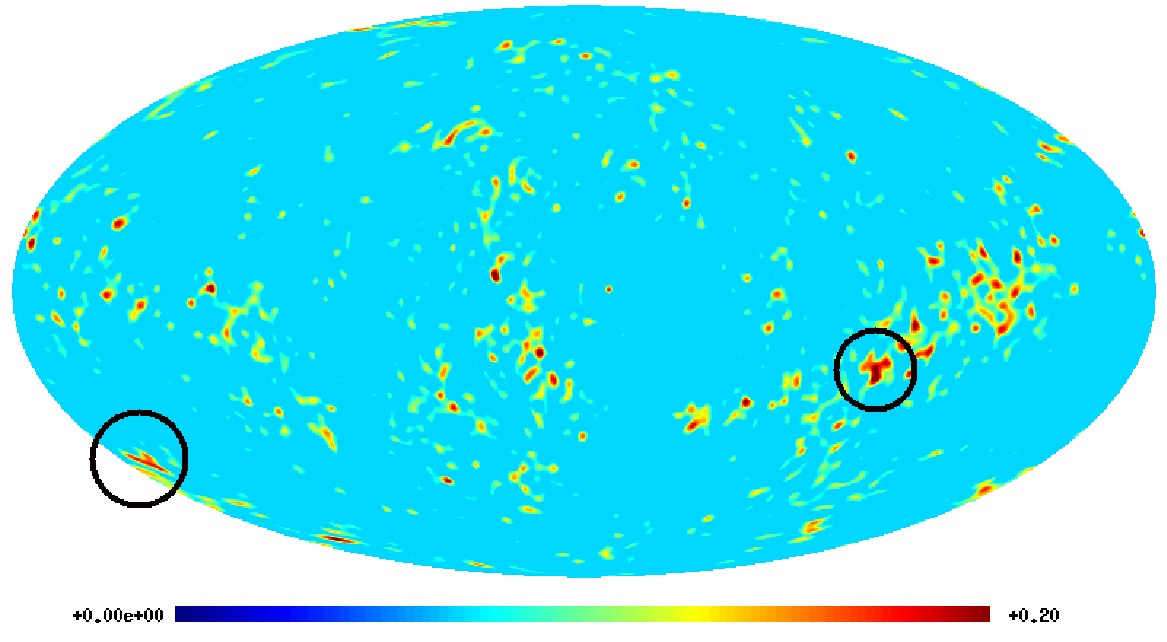}
\caption{The map for
negative (top) and positive (bottom) thresholds $-0.2\le  T\le -0.05 $ and
$0.05\le T\le 0.2 $ of the ILC III  map. For the top map the black
circle marks the location of the CS in the Galactic coordinates. For
the bottom map the circles mark high amplitude positive peaks.
}
\label{f1}
\end{figure}

As was pointed out in the Introduction, historically the CS was
detected in the WMAP data as one of the deepest minima of the CMB
signal, using method based on wavelets. Our first aim is to show
that this CS can actually be detected quite straightforwardly in the
pixel domain using simple threshold techniques. In Fig.\ref{f1}
we take two thresholds of the ILC III signal at the range of
temperatures $-0.2\le \Delta T\le -0.05 $ and $0.05\le \Delta T\le
0.11$\,mK and map them with a color scale $-0.2,0$ and $0,0.2$,
respectively.

\begin{figure}
\setcaptionmargin{5mm}
\onelinecaptionstrue
\centering
\vspace{0.01cm}\hspace{-0.1cm}\epsfxsize=16cm
\epsfxsize=0.45\columnwidth \epsfbox{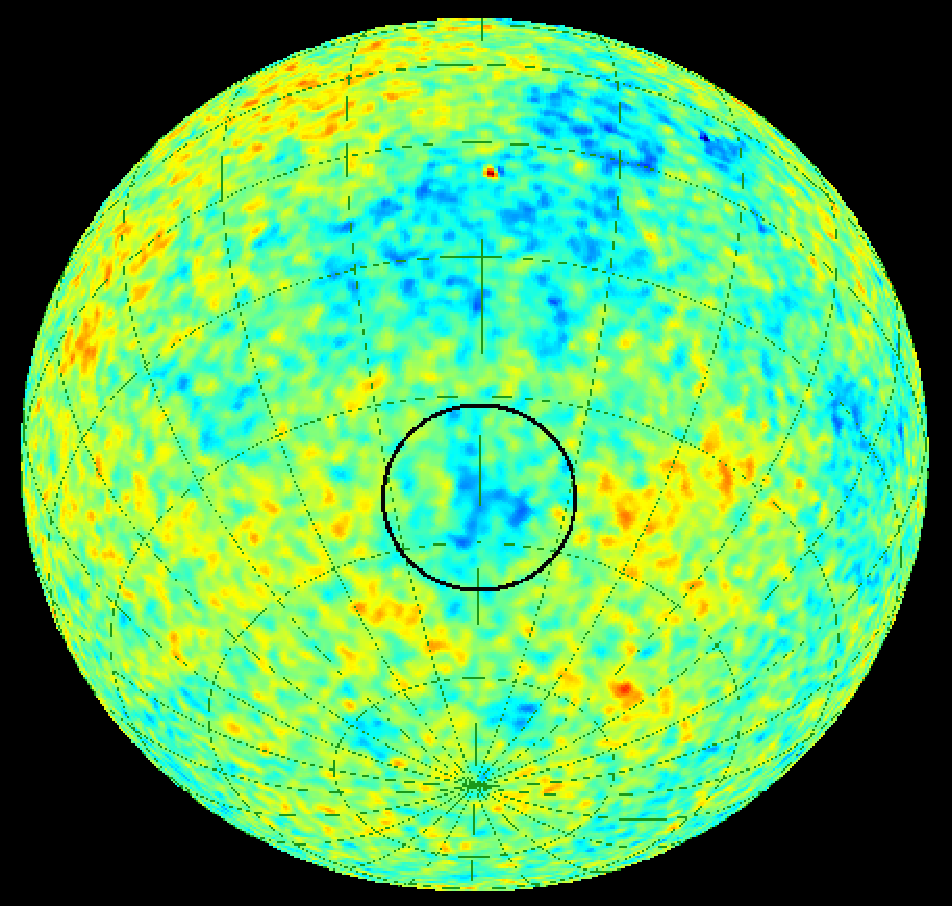}
\centering
\vspace{0.01cm}\hspace{-0.1cm}\epsfxsize=16cm
\epsfxsize=0.45\columnwidth \epsfbox{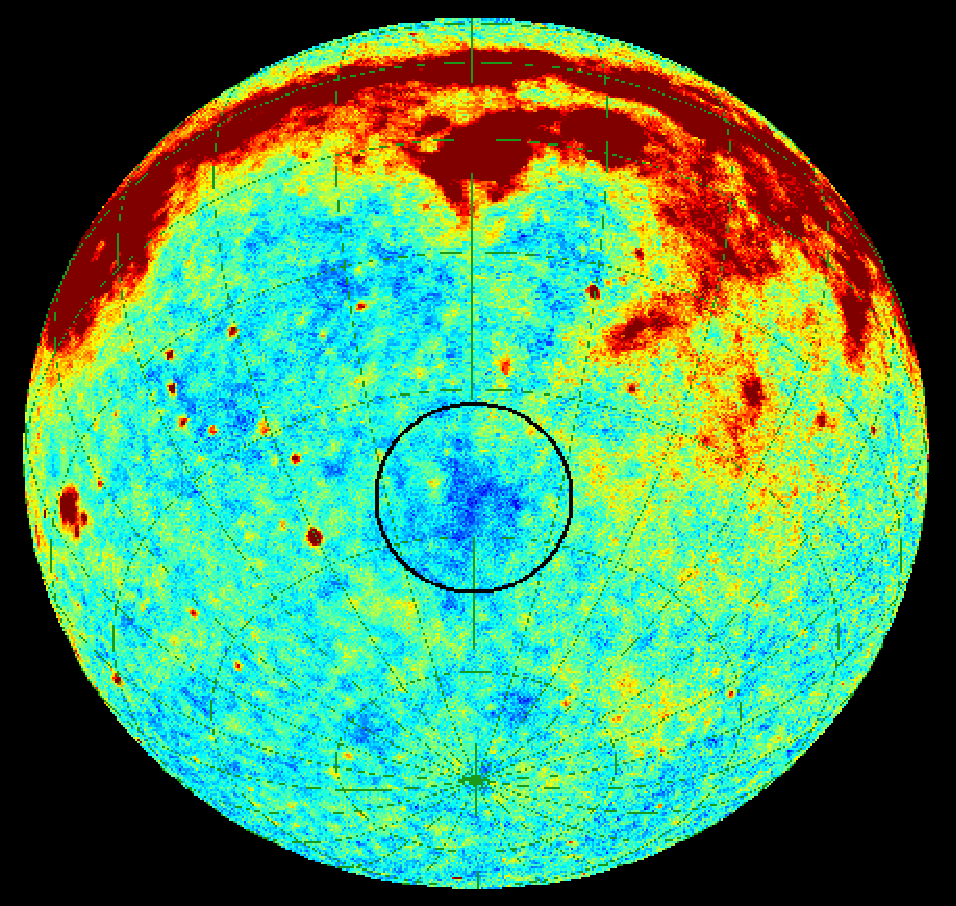}
\centering
\vspace{0.01cm}\hspace{-0.1cm}\epsfxsize=16cm
\epsfxsize=0.45\columnwidth \epsfbox{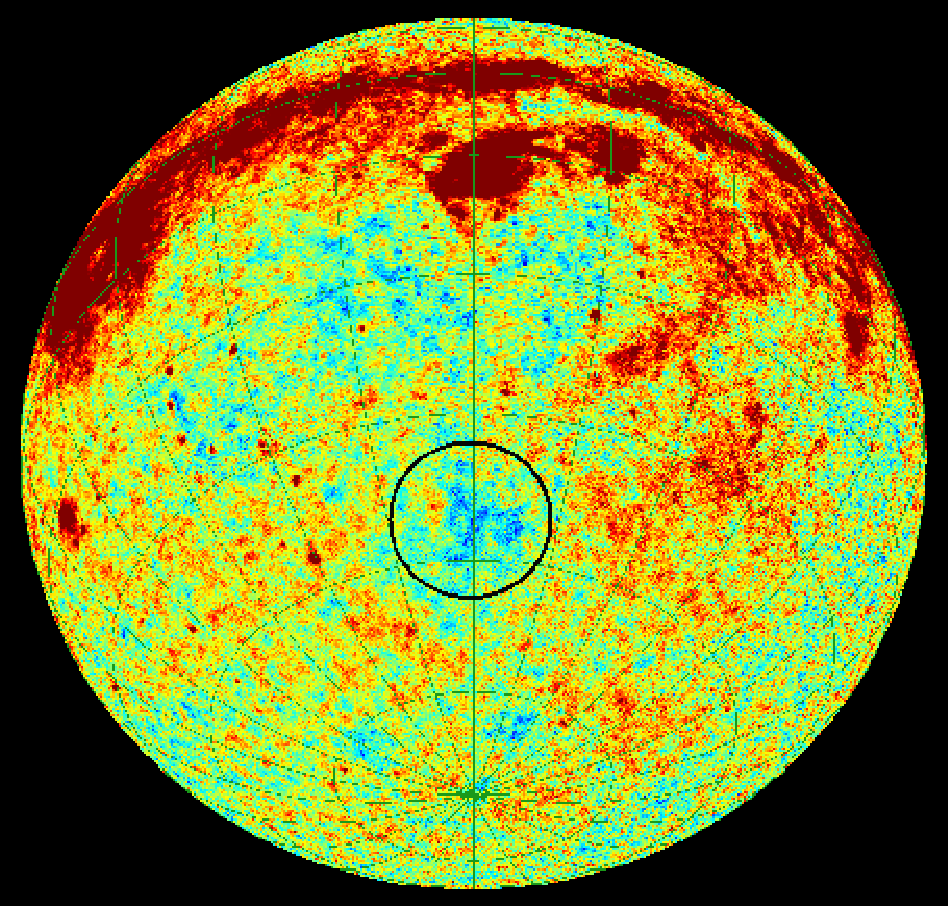}
\caption{The SKY Viewer maps for the ILC III (top),
the K  and Ka bands of the WMAP (second and third from the top)
map with cutoff of intensity down to the minimal values.
For all the  maps the zone of the CS has been placed at the center
of the black circle.
}
\label{f2}
\end{figure}

From this figure one can immediately see the difference in
statistical properties of the positive and negative thresholds of
the map. First of all, it is  important to note that  a majority of
the positive peaks are associated with the  area around $|b|\le
25-30^\circ$ (see the bottom plot), while the negative peaks are
present well outside this region of the map. Secondly, the amplitude
of the highest positive peaks is limited by $0.11$\,mK, while for the
negative peaks it is about  $-0.019$\,mK. For the CS the temperature
of the negative peak is in agreement with the estimate
\cite{cruz2007a}.     
In Fig.\ref{f2}, by using the SKY Viewer,  we map the
ILC III K and Ka maps, in which the CS is  located in the center of
the area marked by the black circle (the blue cluster of peaks
inside the black circle). From the K and Ka maps one can clearly see
that the CS is clearly observed even without subtraction of the CMB
signal. In agreement with
\cite{cruz2005,rudnick} 
we note that there is a cluster of negative
peaks, rather than one single peak.

\begin{figure}
\setcaptionmargin{5mm}
\onelinecaptionstrue
\vspace{0.051cm}\hspace{0.1cm}\epsfxsize=16cm
\centering
\vspace{0.05cm}\hspace{0.1cm}\epsfxsize=16cm
\epsfxsize=0.36\columnwidth \epsfbox{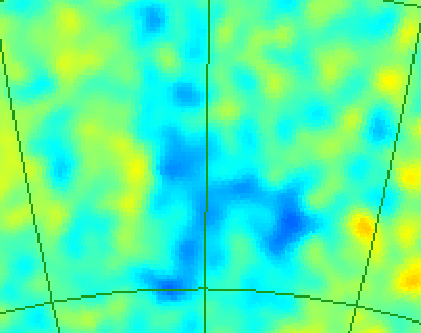}
\centering
\vspace{0.1cm}\hspace{0.1cm}\epsfxsize=16cm
\epsfxsize=0.36\columnwidth \epsfbox{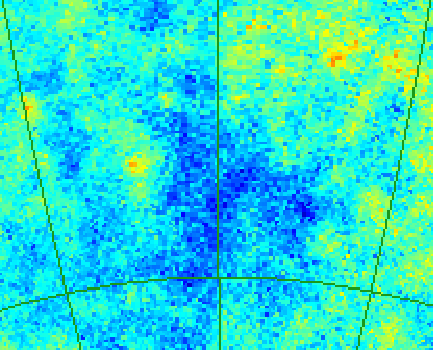}
\centering
\vspace{0.05cm}\hspace{0.1cm}\epsfxsize=16cm
\epsfxsize=0.36\columnwidth \epsfbox{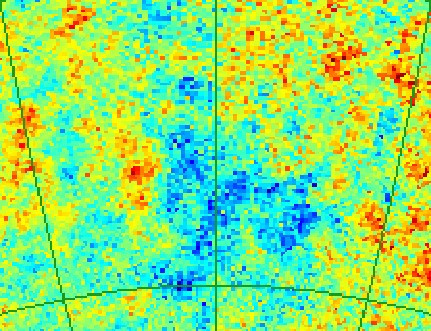}
\centering
\vspace{0.05cm}\hspace{0.1cm}\epsfxsize=16cm
\epsfxsize=0.36\columnwidth \epsfbox{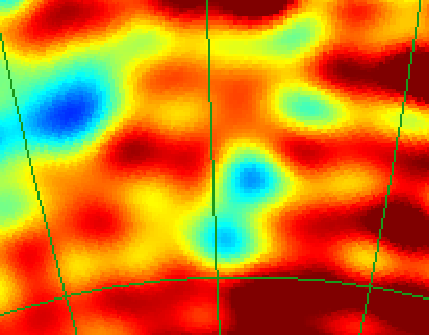}
\caption{The CS zone image projection (in the pixel domain) of the ILC III,
K, Ka and Haslam et al. maps.        
The size of all the maps is $25\times25^\circ$. } \label{f3}
\end{figure}

To show the local structure of the zone containing the CS, in
Fig.\ref{f3} we plot high resolution ($\ell_{max}=100$) images of
the inner structure of the CS zone,
including the Haslam et al. map
\cite{haslam}. 
which is expected to be free from CMB ``contamination''. 

From Fig.\ref{f2} and Fig.\ref{f3} one can draw the important
conclusion that the zone of the CS is surrounded by zones of hot
spots, clearly seen in Fig.\ref{f2} just on the right and left hand
side of the CS.

If the origin of the CS is related to large-angular modulation of
the CMB map and possible anisotropy of the power distribution across
the sky, it would be naturally to expect that additionally to the
WMAP CS detected in
\cite{santanderng}   
 and others, 
we could find other cold and hot spots similar or even equal
morphological properties. To show that this is the case for WCM and
the ILC III maps, in the next section we will use a cluster analysis
of these maps in order to mark possible zones of peculiar
distribution (clusters) in  the signal.

\section{Cluster analysis of the CMB maps in the vicinity of the CS}

The main idea behind our implementation of  cluster analysis  is to
divide the CMB sky above and below the Kp0 mask into iso-latitude
rings and then to analyze the properties of the signal for each ring
separately. First, let us briefly describe the statistical
properties of the signal $T(\theta=\theta_c,\varphi)$ for fixed
latitude $b$.
 Let us take under consideration the distribution of peaks above and
below some given threshold $\nu_t=\Delta T/\sigma_0$, where
$\sigma^2_0$ is the variance for each
ring. For a one-dimensional cross-section of the CMB map we
introduce a definition of a cluster of maxima considering, for
example, two points, $x_1$ and $x_2$ ($x_1<x_2$) from the ring.
If for all the
points in the interval $x_1\le x\le x_2$ we have  $\Delta
T(x)>\nu_t\sigma_0$, we call these points $x_1,x_2$  connected to
each other with respect to the threshold $\nu_t$. A collection of
maxima of $\Delta T$ located at the points $\{x_k\},k=1,2..K$ we
will call a cluster of length $D$ if all the points $\{x_k\}$ are
connected to each other when the threshold $\nu_t$ is applied
\footnote{The same definition can be introduced for cluster of
minima, when instead of maxima we will use the minima of $T$.}. The
length of the cluster $D=|x_1-x_2|/2\pi$ is an analogue of the definition
of the two-dimensional area
\cite{cruz2007a}, 
but in a
one-dimensional case. For a random Gaussian field the statistic of
length of the clusters (one-dimensional) and the area
(two-dimensional) are similarly sensitive to the spectral parameters
of the random field, as described by
\cite{bbks}.   
However, the difference between  cluster analysis and area
statistics becomes obvious in application to non-Gaussian random
fields. Let us assume for a moment that some zone of the signal is
characterized by the area statistic, and that this has a very
non-Gaussian value. To characterize the properties of the
non-Gaussian field it seems to be very important to know, if this
area is related to a uniform structure with one single peak, or
formed from a cluster of peaks with the same area. This is why for
our analysis of the WCM and the ILC III statistical properties we
prefer to use cluster analysis (CA).
In our analysis, in addition to the length of cluster $D$, we will use
its size $S$ which we define as the number of maxima (or minima)
above (or below) the threshold $\nu_t$ within the interval length
$D=|x_1-x_2|/2\pi$.

\subsection{Statistical properties of the signals for equal latitude ring}

To describe the statistical properties of  equal latitude rings we
will use the approach proposed in
\cite{cn}. 

The standard treatment for a full-sky CMB signal $T(\theta,\varphi)$ is
via spherical harmonic decomposition:
\begin{eqnarray}
T(\theta,\varphi)=\sum_{\l=0}^{\lmax} \sum_{m=-\l}^{\l}
	\alm Y_{\lm}(\theta,\varphi),
\label{eq2}
\end{eqnarray}
where $\theta$ and $\varphi$ are the polar and azimuthal angle,
respectively, and $\alm$ are the spherical harmonic coefficients.
Here $\ylm$ are the spherical harmonics, defined in terms of
Legendre polynomials and plane waves:
\begin{equation}
\ylm(\theta,\varphi)=N_\lm P^m_\l(\cos \theta) \exp(i m \varphi),
\label{eq2}
\end{equation}
where
\begin{equation}
N_\lm=(-1)^m \sqrt{\frac{(2\l+1)(\l-m)!}{4\pi(\l+m)!}}.
\label{eq4}
\end{equation}
 For a random Gaussian CMB sky the properties of the statistical ensemble
of realizations  are completely specify by the power spectrum
\begin{equation}
C(\l)=\frac{1}{2\l+1}\sum_{m=-\l}^{m=\l} \langle|a_{\l,m}|^2\rangle,
\label{eq5}
\end{equation}
while for each single realization we expect to find some deviation
from $C(\l)$ due to the ``cosmic variance'' effect. Below we will
use a polar coordinate system in which the Galactic plane ($b=0$) is
associated with $\theta=\pi/2$.

Let us analyze the signal $T(\theta_c,\varphi)$  from the
equal-latitude ring at  $\theta=\theta_c$, where $|\theta_c|\ge
|\theta_{\rm mask}|$, where $|\theta_{\rm mask}|$ is the maximum
latitude of any foreground masks. This ring
$T(\theta_c,\varphi)\equiv T_c(\varphi)$ is a one-dimensional
signal, for which we can construct a Fourier transform  with
coefficients $g_m$:
\begin{equation}
T_c(\varphi)=\sum_{m=-\lmax}^{\lmax} g_m \,\exp(i m \varphi),
\label{eq6}
\end{equation}
where
\begin{equation}
g_m=\int_0^{2\pi} d\varphi \, T_c(\varphi)\, \exp(-im\varphi).
\end{equation}
We can then relate the ring to the full-sky signal via
Eq.(\ref{eq2}) and (\ref{eq6}) and get
\begin{equation}
g_m=\sum_{\l\ge |m|}^{\lmax} N_\lm\, P_\l^m(\cos\theta_c)\, \alm.
\label{eq7}
\end{equation}
That is to say that the Fourier coefficients $g_m$ of the ring can
be expressed as a combination of the full-sky $\alm$. Defining the
variance of the signal for equal latitude ring as
\begin{equation}
Var T=\frac{1}{2\pi}\int_0^{2\pi}d\varphi \left[T(\varphi)-
       \langle T \rangle \right]^2,\hspace{0.05cm}
\langle T\rangle=\frac{1}{2\pi}\int_0^{2\pi}d\varphi T(\varphi)
\label{eq8}
\end{equation}
and after substitution of Eq.(\ref{eq6}) into  Eq.(\ref{eq8}) we have
\begin{eqnarray}
\langle T\rangle=\sum_{\l}N_{\l,m=0}P_\l(\cos\theta)a_{\l,m=0},\nonumber\\
Var T=\sum_\l\sum_{\l'}\sum_{m\neq 0}N_{\l,m}N_{\l',m}P_{\l,m}
    (\cos\theta)\times\nonumber\\
\times P_{\l',m}(\cos\theta)a_{\l,m}a^{*}_{\l',m}. \label{eq9}
\end{eqnarray}

For a random Gaussian field (GRF), after average over realizations,
the combinations of the $\alm$ coefficients in Eq.(\ref{eq9})
satisfy the following conditions:
\begin{equation}
\langle a_{\l,m}a^{*}_{\l',m'}\rangle = C(\l)\delta_{\l,\l'}\delta_{m,m'}
\label{eq10}
\end{equation}
 and then
\begin{eqnarray}
Var T=\sum_\l\sum_{m\neq 0}N^2_{\l,m}P^2_{\l,m}(\cos\theta)
       C(\ell)=\sum_\l W(\l,\cos\theta)C(l)\nonumber\,,\\
\label{eq11}
\end{eqnarray}
where $W(\l,\cos\theta)=\sum_{m\neq 0}N^2_{\l,m}P^2_{\l,m}(\cos\theta)$
is the window function of the ring.

\begin{figure}
\setcaptionmargin{5mm}
\onelinecaptionstrue
\centering
\vspace{0.01cm}\hspace{-0.01cm}\epsfxsize=16cm
\epsfxsize=1.\columnwidth \epsfbox{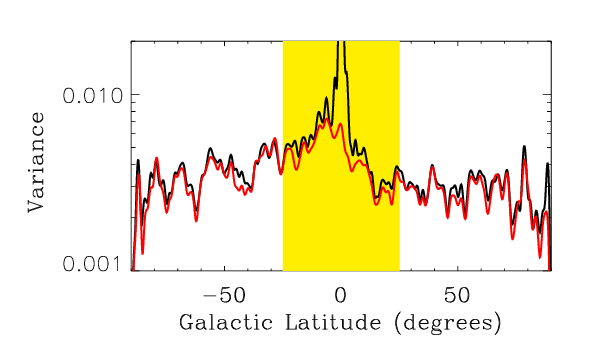}
\caption{The variance for the WCM map (the black line) and
the ILC III map (the red line) versus the galactic latitude.
}
\label{ff}
\end{figure}

 Our approach here is a special case for a well known theoretical prediction:
 any $n$ dimensional cross sections of $N$ dimensional Gaussian random signal
produce a Gaussian signal as well. However, this general theory
tells us about the
GRF, represented as a sets of realizations. For one single
realization of the CMB sky, as the WMAP signal does, the
Eq.(\ref{eq11}) is no longer available. More general, instead of
Eq.(\ref{eq11}) we have
\begin{equation}
a_{\l,m}a^{*}_{\l',m'}=C(\l)G^{m,m'}_{\l,\l'},
\hspace{0.1cm}\langle G^{m,m'}_{\l,\l'}\rangle=\delta_{\l,\l'}\delta_{m,m'}\,.
\label{eq12}
\end{equation}
Thus,
\begin{eqnarray}
Var T=\sum_\l\sum_{\l'}\sum_{m,m'\neq 0}N_{\l,m}N_{\l',m'}P_{\l,m}(\cos\theta)
	\times\nonumber\\
P_{\l',m'}(\cos\theta)C(l)G^{m,m'}_{\l,\l'},\nonumber\\
W(\l,\cos\theta)=\sum_{\l'}\sum_{m\neq 0}N_{\l,m}N_{\l',m'}\times\nonumber\\
P_{\l,m}(\cos\theta)P_{\l',m'}(\cos\theta)G^{m,m'}_{\l,\l'}
\label{eq14}
\end{eqnarray}
The matrix $G^{m,m'}_{\l,\l'}$ describes  the coupling between different
modes $\l,m$ and $\l',m'$ for the random process
$T(\varphi)$, which leads to variations of the variance $Var T(\theta)$
for different rings ($\theta=const$).
One can see that integration of Eq.(\ref{eq14}) over $\theta$ preserves
spontaneous correlations
($G^{m,m'}_{\l,\l'}\neq \delta_{\l,\l'}\delta_{m,m'}$)
even for the whole sky. However, for a particular  ring we will have
an additional modulation of these correlations depending on the $\theta$
through the window function $W(\l,\cos\theta)$.
In Fig.\ref{ff} we show the dependence of $Var T$ on the galactic latitude
for the WCM and the ILC III maps.

First of all, we stress the remarkable similarity of these maps
outside the region of the Kp0 mask. Moreover, the variance of the
WCM map per each ring matches the variance of the ILC III almost
exactly for all latitudes, except for the  $|b|\le 5^\circ$ zone.
Secondly, note that the $b=-57^\circ$ ring lies near to the local
maxima of the variance. The width of this zone is about $\theta\sim
10^\circ$. Thirdly, from Fig.\ref{ff} one can clearly see the
asymmetry of the variance for the rings located at the North and the
South hemispheres.
To characterize this asymmetry we introduce the following parameter
\begin{eqnarray}
A(x=\cos\theta)=\frac{Var T (-\cos\theta)-Var T(\cos\theta)}{Var T (\cos\theta)+Var T(-\cos\theta)}
\label{eq19}
\end{eqnarray}
where $Var T(\cos\theta)$ is the variance of the signal for the ring
with polar coordinate $\theta$.
\\
\begin{figure}
\setcaptionmargin{5mm}
\onelinecaptionstrue
\centering
\vspace{0.01cm}\hspace{-0.01cm}\epsfxsize=16cm
\epsfxsize=0.9\columnwidth \epsfbox{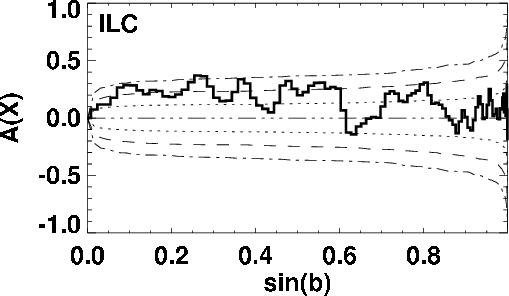}%
\caption{The asymmetry parameter $A(x)$ versus $\sin(b)$ for the ILC\,III
rings and for 1000 realizations of the
RGF CMB. The dotted , dashed and dash-dotted lines correspond
to 68$\%$, $95\%$ and $99\%$ threshoulds.
}
\label{ffa}
\end{figure}

In Fig.\ref{ffa} we show the behavior of the parameter of asymmetry
$A(x)$ versus $\sin(b)$ in comparison with the same parameters taking
from 1000 realizations of the RGF CMB.
 No additional  analysis is required to observe that
the variance  of the Southern hemisphere is higher than for the
North. This result is in agreement with
investigations of
the asymmetry of the power per $10^\circ$ patches of the sky
\cite{eriksenasym},
and
analysis of peak statistics
\cite{larson}.
Furthermore, looking at Fig.\ref{ff} at
$b=-25^\circ$ one can find the global maxima of the variance outside
the Kp0 mask. This zone is clearly seen in Fig.\ref{f1} just below
the galactic plane as a large cluster of minima mentioned by Park
\cite{park}  
Could the origin of
this cluster be the same as for the cluster around the CS~? To
answer this question we need to look closely on the properties of
the signal for the ring $b=-57^\circ$ in the azimuthal direction.
Note that this approach is close to the Eriksen et al. method
\cite{eriksenasym},
when the whole  CMB sky
was divided in  zones with characteristic scales (diameter)
about $10^\circ$, and after that
for each zone the power of the signal was in used to characterize
the difference between them.
In our approach, we partially use the Eriksen et al. method,
dividing the CMB sky in iso-latitude
circles, and then by analysis of the morphology of the signal for each
ring we will try to find a possible
source of peculiarity of the signal. From Fig.\ref{ffa} clearly seen
that the rings at $|\sin(b)|> 0.82$ are characterized by the values of
the parameter of asymmetry $A(x)\le 0.2$.
However, small values of this parameter
tell us that the distribution of variance for each ring
versus the galactic latitude is nearly symmetrical in respect to
the Galactic plane.
Thus,  the CS  detected for the $b=-57^\circ$,
could  have a  ``mirror partner'' for the ring at $57^\circ$,
or at least, the morphology of the signal for these rings could
be close to each other to provide nearly the same variance
$VarT(b=-57^\circ)\simeq VarT(b=57^\circ)$, and $A(x)\le 0.2$.

\subsection{The clusters in the ring at $b=-57^\circ$}

\begin{figure}
\setcaptionmargin{5mm}
\onelinecaptionstrue
\centering
\vspace{0.01cm}\hspace{-0.01cm}\epsfxsize=16cm
\epsfxsize=0.8\columnwidth \epsfbox{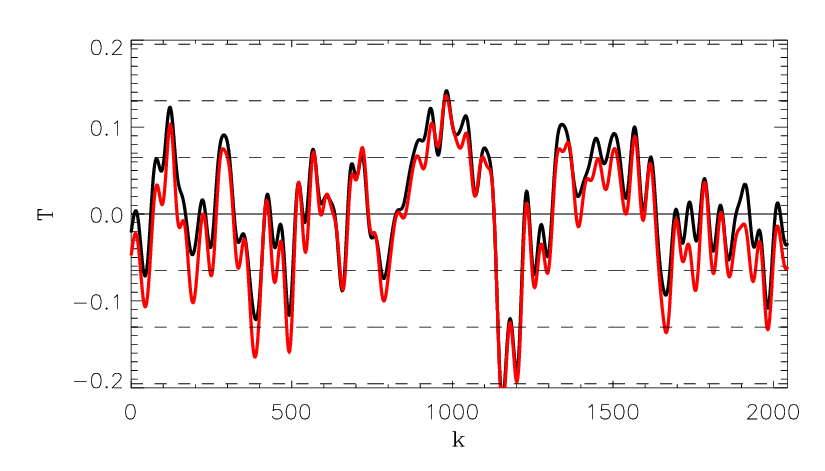}
\caption{ Plot of the temperature distribution in
the one-dimensional ring through the center of the
CS ($b=-57^\circ$) in azimuthal direction. The solid
line
corresponds to the WCM signal  and the red line is for the ILC III
map. All the maps were smoothed by the angle $\Theta_c=1^\circ$. For
this ring $\sigma_0=0.0651$\,mK. The dashed lines mark the thresholds
$|\nu_t|=1,2,3$. }
\label{f4}
\end{figure}

In this section we draw attention to the azimuthal distribution of
the signal for the iso-latitude ring with $b=-57^\circ$, which
contains the CS.   In Fig.\ref{f4} we plot $ T(\theta_c, \phi)$  for
WCM and the ILC III maps smoothed by the angle $\Theta_c=1^\circ$.
This figure clearly demonstrates that there are no significant
differences  of morphology of the CS in the ILC III and WCM maps
Note that the WMAP team have pointed out that the ILCIII map is suitable
 for scientific analysis only for the range of multipole momentum $\l\le 10$.
 As it is seen from Fig.\ref{f4} , for $b=-57^\circ$ ring there are no
significant differences between the ILC III and WCM maps. This tendency
is stable and for other rings outside the Kp0 mask.
This is why below we use the
name of the ILC III map as an indicator. All the analysis presented
in the paper
was performed  for the WMAP ILC III maps in combination with the WCM map, and
we did not find any significant differences between  them.

In
the following analysis we change the reference system of coordinate
from the Galactic one to one in which $\phi=0$ is associated with
the Galactic center and then all the values of $\phi$ are counted
clockwise up to $\phi=360^\circ$. The GLESP pixelization allow us to
fix the same number of pixels for each iso-latitude ring ($N=2048$)
and consequently the location of each pixel $0\le k\le N$ is related
to the angle  $\phi_k$ as $\phi_k=2\pi k/N$.

\begin{figure}
\setcaptionmargin{5mm}
\onelinecaptionstrue
\centering
\vspace{0.01cm}\hspace{-0.1cm}\epsfxsize=16cm
\epsfxsize=0.8\columnwidth \epsfbox{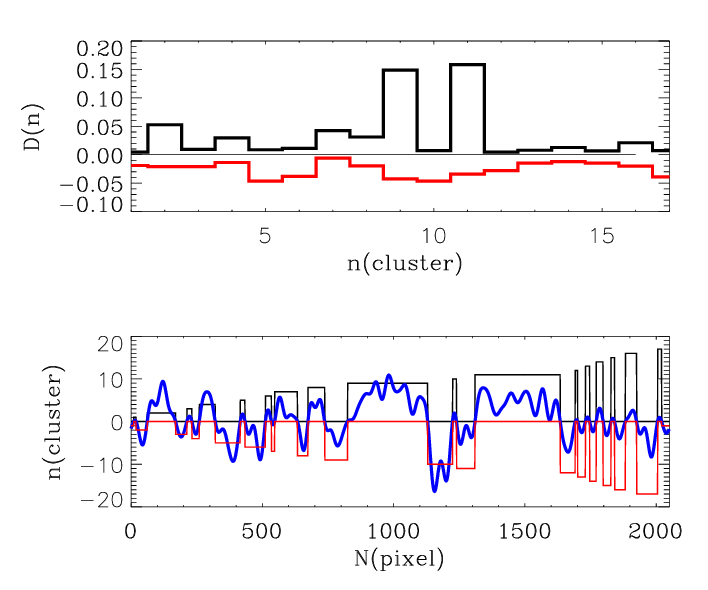}
\caption{Top.
The length $D(n)$ of the clusters
$\nu_t=0$ versus the number of cluster. The
positive
line
is for clusters of maxima, the
negative
line represents the minima.
Bottom plot. Number of cluster versus its position in the pixel
domain (2048 pixels correspond to the $360^\circ$ in azimuthal
direction).
The
thick
line corresponds to a one--dimensional ring of
the ILC III map trough the CS  in units of $5* T_k/\sigma_0$. The
negative sign for $D^{-}(n)$ marks the length of the clusters of
minima. }
\label{f5}
\end{figure}

In Fig.\ref{f4} the CS is clearly seen at $k\simeq 1130 \div1224$ as
a cluster of  minima (C10, n=10) of the signal in the WCM and ILC
III maps. Moreover, there are no significant differences between
these two signals either for the whole ring, or for the zone of the
CS.  Fig.\ref{f5} shows the distribution of the length of the
clusters versus the number of cluster. Clusters of maxima are shown
in black, minima in red. As one can see from Fig.\ref{f5} there are
about 17 clusters of maxima and 17 clusters of minima. The mean
length of  the ILC and WCM clusters of maxima $\langle D^{+}_{\rm
ilc}(n)\rangle \simeq 0.0337$, while for the clusters of minima we
get $\langle D^{-}_{\rm ilc}(n)\rangle\simeq 0.0256$. For
comparison, after taking 1000 realizations for the GRF with the WMAP
best fit $\Lambda$CDM power spectrum  we have found $\langle
D^{+}_{\rm grf}(n)\rangle\simeq 0.0303$ and $\langle D^{-}_{\rm
grf}(n)\rangle\simeq 0.0315$. Remember that in the
one-dimensional case the area statistic, used in
\cite{cruz2007a}    
for
two dimensional case and the length statistic, applied above are
equivalent. Nevertheless, to estimate the probability distribution
function $P(D)$ for the clusters at $\nu_t=0$ and to have a length $D$
within the interval $D-\delta D\div D+\delta D$ we performed the
numerical simulations using 1000 realizations of the RGF CMB. In
particular, for each of realization  we took into account that the
length of each ring with corresponding latitude $\theta$ is rescaled
by factor $|\sin\theta|$, which re-normalize the mean length of the
cluster for each ring. Then we got the number of the clusters with
length $D$ versus its length and come to the distribution function
$P(D)$. For $D>\langle D\rangle$ the function $P(D/\langle
D\rangle)$ has Gaussian form $P(D)\propto \exp(-0.5(D/\langle
D\rangle)^2)$ and corresponding probability to get the cluster with
the length $D/\langle D\rangle\simeq 3-4$ is about $0.00033-0.011$.
For the ring with $\theta =\pi/2$ the total number of clusters is
about $40$ (for chosen resolution of the map $\Theta=1^\circ$), and
the existence of a single cluster with $ D/\langle D\rangle\sim 3$
is quite possible. However, the probability to get two or even three
clusters with $D/\langle D\rangle> 3$ for one ring is less then
$0.016$ ( for two clusters with $D/\langle D\rangle= 3$), and less
then $0.005$ for three clusters with the same length. For $D/\langle
D\rangle> 4$ the corresponding probability to find two clusters is
less then $2\cdot10^{-4}$. Note that our simulations are based on the RGF
CMB map, which generally reproduced the statistically homogeneous
and isotropic Gaussian random process. As it was pointed out in
Section 3.1, the ILC III map is already statistically peculiar.
Therefore, the model of the RGF CMB is not quite adequate to the
WMAP ILC III or WCM maps, and corresponding estimation of the
statistical properties of these signals needs to be taking into
account as a tendency of departure from Gaussianity, rather then
strong evidence of it. In Section 4 we will discuss some of the
possible consequences of the anisotropy of the power , discussed in
Section 3.1 and their influence on the properties of the zone with
the CS.

Coming back to the analysis of the CS, we would like to mention that the
 cluster C10 containing the CS has length $D_{\rm CS}=0.0459
\simeq 1.5 \langle D^{-}_{\rm grf}(n)\rangle $ and its dimension is
$D^{-}=2$ at $\nu_t=0$. At the same time $D_{\rm CS}/\langle
D^{-}_{\rm ilc}(n)\rangle\simeq 1.8$ at $\nu_t=0$. These ratios tell
us that this cluster looks like an ``ordinary'' cluster defined by
the threshold $\nu_t=0$ without any significant deviation from the
statistical properties of a random Gaussian Field. However, the
extraordinary properties of this cluster become obvious when we take
into consideration other thresholds, $-3\le \nu_t\le -2$. From
Fig.\,\ref{f5} we see that even for this range of $\nu_t$ the
cluster still appears as a cluster of minima at the level $\nu_t=-3$,
and local maxima at $\nu_t=-2$.
Novikov and Jorgensen 
\cite{novjorg}
 have pointed out that for the cluster  of
minima  with $d^{-}=2$ the conditional probability $P(k_1, k_2)$ to
find two negative peaks located at the points $k_1$ and $k_2$ in
this cluster is
 less than   the conditional probability to find one maximum in between
these points with height above $\nu_t$. In our case this threshold
is $\nu_t\simeq-2$ and $P(k_1, k_2)\le 0.02$.  Note that the
conditional probability $P(k_1, k_2)$ had been found theoretically
in
\cite{novjorg}
for a random Gaussian field. In
practice to prove this expectation we took under consideration 1000
realizations of the whole sky CMB maps from the random Gaussian
generator and fix only those of the points of extrema (minima and
maxima) which exceed the threshold $2\sigma$. Then, for these points
of extrema, we select the only those which have a structure of a
cluster with two maxima and the single minimum between them (or two
minima and the local maximum between them) with the amplitude of  the
corresponding minimum (or maximum) $\nu_{m}\ge 2\sigma$, where
$\sigma^2$ is the variance of the CMB signal for each realization of
the CMB sky. The angular resolution of the RGF CMB maps corresponds
to the $1^\circ$ and the power spectrum of the CMB $C(\ell)$ used for
generation of the ensemble of realizations corresponds to the WMAP
best fit $\Lambda$CDM cosmological model
\cite{wmap5ycos}.
We have found that 38 (17 minima and 21 maxima) realizations from
1000 reveal similar to the CS morphology for particular isolated
zones. However, none of those 38 realization were surrounded by the
positive (or negative) clusters with dimension 4 or 5. After
detection of the isolated zones similar in morphology to the CS, we
extracted one-dimensional cross-sections of the map through these
zones in order to compare the properties of the one dimensional scan
and the $b=-57^\circ$  ring from the ILC III map. Neither
instrumental noise nor the beam profile were taken into account. Our
motivation for using this simplest model is that the characteristic
scale of the CS, mentioned in
\cite{cruz2005,cruz2007a}
is about
$10^\circ$ . For these angular scales the contribution of the
instrumental noise and the antenna beam shape are too small to
affect the properties of the signal in the vicinity of the CS.
Moreover, the structure of the CS is practically the same as for the
WMAP first, third and fifth year data release, for which the
instrumental noise is significantly different.  We are planning to
generalize our analysis, including mentioned above instrumental
noise and beam shape in a separate paper.

Coming back to the analysis of the  CS, we would like to point out
that the CS  cluster is not a unique  feature of the $b=-57^\circ$
ring. There are two  clusters of maxima (see  Fig.\ref{f4}), namely
the cluster C9 with $k_{\rm min}=825$ and $k_{\rm max}=1129$ with
$D_{C9} =0.15= 4.40\langle D^{+}_{ilc}(n)\rangle $ and the cluster
C11 with $k_{\rm min}=  1310, k_{\rm max}= 1634$, and $D_{C11}
=0.15= 4.69\langle D^{+}_{ilc}(n)\rangle $. Here $k_{\rm min},k_{\rm
max}$ mark the coordinates of the cluster at $\nu_t=0$ in the pixel
domain. These clusters seen in Fig.\ref{f5} are near to the cluster
of minima C10 within which the CS lies.

\begin{figure}
\setcaptionmargin{5mm}
\onelinecaptionstrue
\centering
\vspace{0.01cm}\hspace{-0.01cm}\epsfxsize=16cm
\epsfxsize=0.8\columnwidth \epsfbox{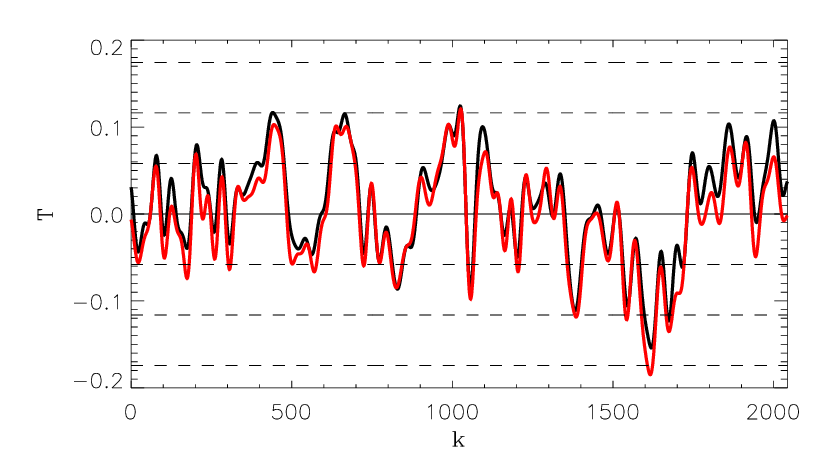}
\epsfxsize=0.8\columnwidth \epsfbox{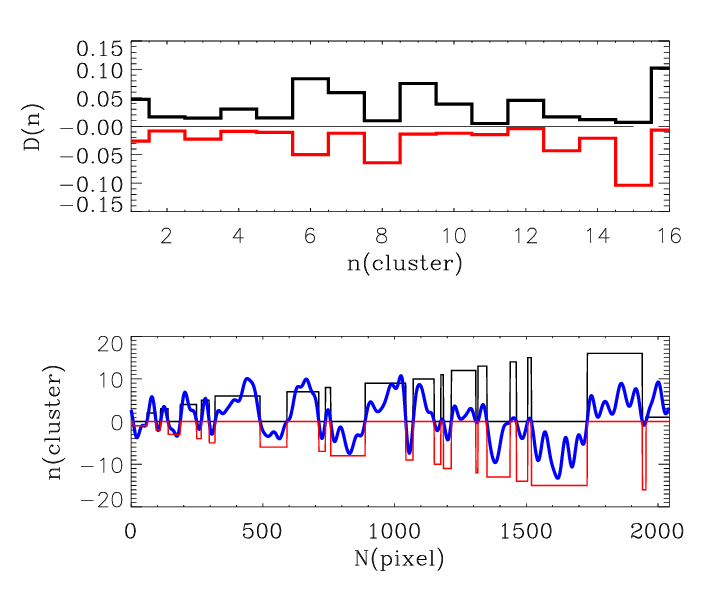}
\caption{The same as for Fig.\ref{f4} and Fig.\ref{f5}, but for the ring
with $b=57^\circ$.  }
\label{ff5a}
\end{figure}

The existence of such clusters with lengths above $4\langle
D^{+}_{\rm ilc}(n)\rangle$ and the cluster of minima C10 with such
large amplitude (negative) peaks seems to be a quite peculiar
feature of the $b=-57^\circ$ ring. However, these peculiarities are
not so specific. To show that ``over-clustering'' of the CMB sky is
a typical feature of the signal let us examine, for example, the
ring with $b=+57^\circ$, located symmetrically to the $b=-57^\circ$
ring in respect to the Galactic plane. In Fig.\ref{ff5a} we show
$T(k)$, and the distribution of clusters versus their length for the
ring with $b=57^\circ$. This signal reveals a remarkable similarity
of the morphology to the ring $b=-57^\circ$. As for the ring with
CS, the ring $b=57^\circ$ is characterized by very high level of
clusterization, the existence of the $3\sigma$ minimum, as a member of
the cluster with $D^{-}=4$, and the existence of the positive
cluster with $D^{+}=6$. Moreover, one can see that the cluster of
maxima C7 at $k_{min}=547,k_{max}=633 $ has a structure  similar to
the structure of the signal in the CS, but now for the maxima.

At the end of this section we would like to point out that the
existence of  clusters with the length $D> 3\div4\langle D\rangle$
is a quite rare event  for the GRF. The presence of three clusters
with $D\sim 3\langle D\rangle$  for single iso-latitude ring at
$b=57^\circ$, and two clusters with $D> 4\langle D\rangle$ for
$b=-57^\circ$, as the CS as well, allow us to conclude  that the ILC
III and WCM maps are generally ``over-clustered''. Attention on the
CS was focused mainly because of the {\em amplitude} of the signal
in that position. A more specific feature of the CS is not this, but
that it is really a cluster of peaks with nearly the same amplitude,
and there is a very large cluster near to it.

\section{``De-clustering'' of the ILC III and WCM by Linear filtration}

In this Section we re-examine an idea of Cruz et al.
\cite{cruz2007a}, 
that the origin of the CS can be explained if the WMAP CMB signal
contains non-Gaussian components in combination with  the Gaussian
ones. This idea seems to be quite natural  since the low multipole
part of the WMAP ILC III map reveals significant peculiarities of
the signal: alignment between quadrupole and octupole, coupling with
the Galactic foregrounds, low power of the quadrupole, and so on.
Note that the WMAP team performed the analysis of Gaussianity of the
CMB signal by subtracting the ten multipole components of lowest
order from the map and claimed that the rest of the signal is in
agreement with  Gaussian statistics. However, asymmetry of the power
of the CMB, discovered
\cite{eriksenasym}  
the existence of
the CS and another peculiarities, mentioned in
\cite{mcewen2006b,mcewen_nvss},        
raise the question of whether there is
some mark imprint non-Gaussian features of the CMB, mainly localized
in the low multipole range of the power spectrum.

If the WMAP CMB outside the Galactic mask $|b|=25^\circ$ is the sum of
Gaussian and non-Gaussian components, it seems natural to guess that
these two signals would have different characteristic scales. The
idea, which we will develop below,  is to use a linear filter of the
CMB signal for each ring of the map with variable scale of
filtration $R$, which can divide the CMB signal in two parts:
$S(k)=\overline S(k,R)+s(k)$, where $s(k)$ corresponds to Gaussian
component and $G(k)=\overline S(k,R)$ corresponds to non-Gaussian
one. Note that Cruz et al.
\cite{cruz2007}   
use the wavelet approach to
perform this analysis. We will instead use the running window filter
defined in the pixels domain as
 \begin{eqnarray}
\overline S(k,R)=\langle T_k\rangle =
    \frac{1}{R}\sum_{j=0}^{R-1}T_{k+j-\frac{R}{2}},\nonumber\\
    \hspace{0.5cm}j=[\frac{R-1}{2},..N_{pix}-\frac{R+1}{2}]\nonumber\\
    s_k=T_k-\langle T_k(R)\rangle. \nonumber\\
 \label{filt}
 \end{eqnarray}
Note that the choice of the linear filter is not so important for
the criteria of separation of the Gaussian and non-Gaussian tails of
the signal. One can for example use the Gaussian filter $\overline
S(k)\propto \sum_j T_j\exp\left (-(k-j)^2/R^2\right)$, or any
another reasonable filter. What is important is the chose of the
criteria of the scale $R$. Following
\cite{cruz2006}   
we will use the skewness and kurtosis of the signals $s_k$ as a
functions of $R$ trying to minimize the difference between their
actual values and most probable values for the GRF.

\begin{figure}
\setcaptionmargin{5mm}
\onelinecaptionstrue
\centering
\vspace{0.01cm}\hspace{-0.1cm}\epsfxsize=16cm
\epsfxsize=0.7\columnwidth \epsfbox{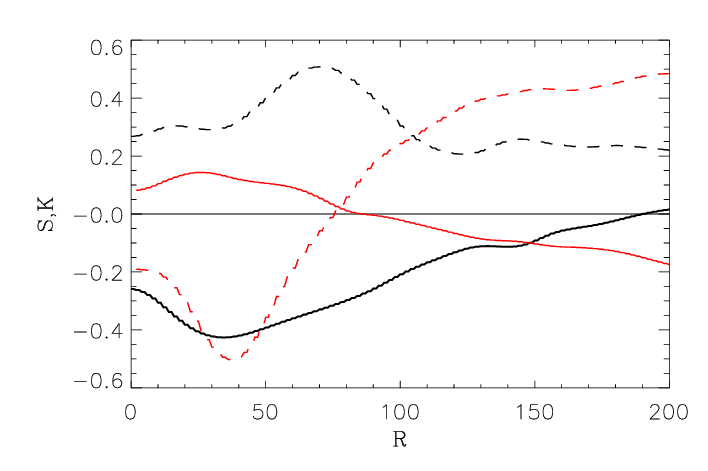}
\epsfxsize=0.7\columnwidth \epsfbox{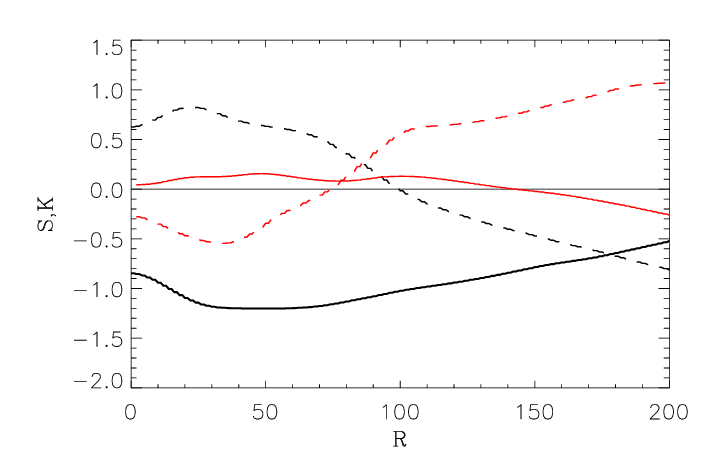}
\caption{Top. The skewness  and kurtosis  of the ILC III $b=-57^\circ$ ring.
The solid
and the dash
lines correspond to the skewness and
kurtosis of the smoothed part of the ILC III signal $\overline S(k,R)$.
The
grey
lines are for the signal $s_k$. Bottom. The same as top, but
for the zone of the ring $k_{min}=825,k_{max}=1634$.
}
\label{skew}
\end{figure}

In Fig.\ref{skew} we show the skewness and kurtosis for the ILC III
ring $b=-57^\circ$ for different scales of filtering $R$.

\begin{figure}
\setcaptionmargin{5mm}
\onelinecaptionstrue
\centering
\vspace{0.01cm}\hspace{-0.1cm}\epsfxsize=16cm
\epsfxsize=0.6\columnwidth \epsfbox{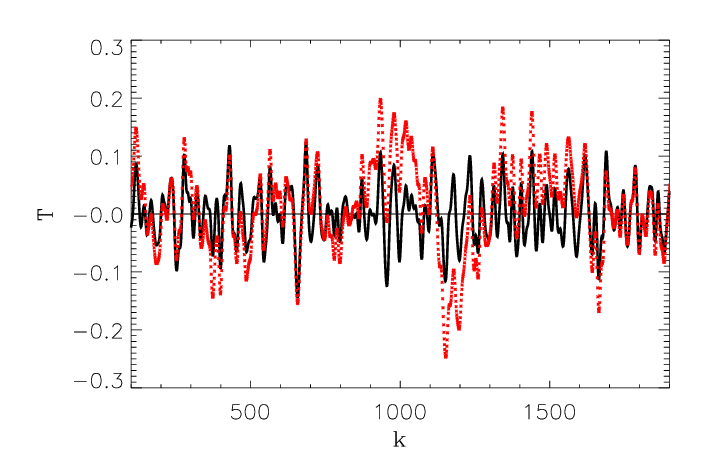}
\epsfxsize=0.6\columnwidth \epsfbox{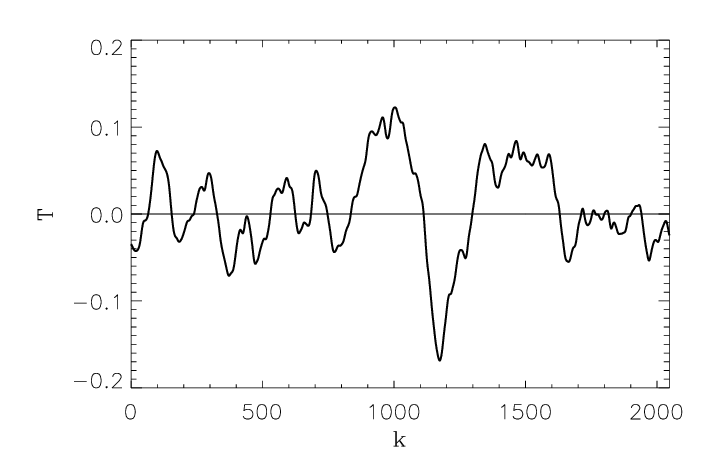}
\epsfxsize=0.6\columnwidth \epsfbox{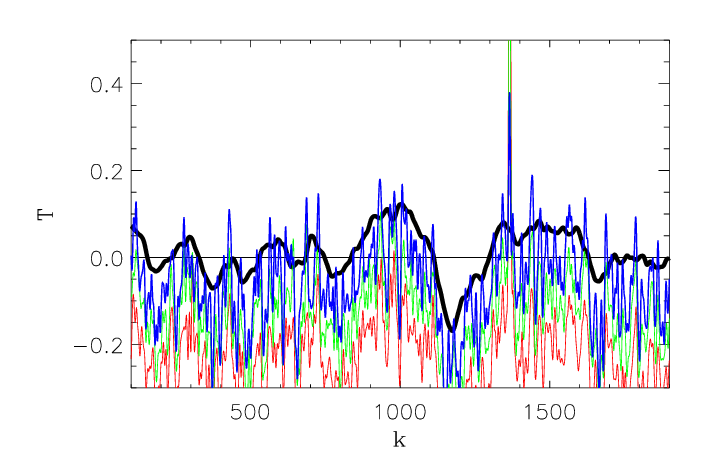}
\caption{Top. The ILC III ring $b=-57^\circ$ before demodulation
(grey dots) and after (black line). Middle. The ILC III signal,
smoothed over $R=70$ pixels (the corresponding angular scale
$\Delta\phi=12.3^\circ$.) Bottom. The smoothed ILC signal (thick
black line) in comparison with signals from Ka,
Q
and V band.
}
\label{dem}
\end{figure}

The skewness and kurtosis are very sensitive to the choice of the
zone of the ring. For example, if we were to look at the zone
$k_{\rm min}=825,k_{\rm max}=1634$ occupied by the clusters
$C^+9,C^-10,C^+11$ we get the skewness and kurtosis shown in
Fig.\ref{skew} (bottom plot). From these two plots one can see that
the value of the parameter $R\simeq 70$ is preferred for this
analysis. For this scale of filtering the kurtosis of the smoothed
signal reaches a maximum, while for the Gaussian tail of the total
signal both the characteristics are close to zero both for the whole
ring and for the particular zone around the CS. In  Fig.\ref{dem} is
shown the ILC III signal before (the red dotted line) and after
subtraction of the smoothed over $R\simeq70$ pixels (the black
line). One  sees that the CS is eliminated.

\begin{figure}
\setcaptionmargin{5mm}
\onelinecaptionstrue
\centering
\vspace{0.01cm}\hspace{-0.1cm}\epsfxsize=16cm
\epsfxsize=0.45\columnwidth \epsfbox{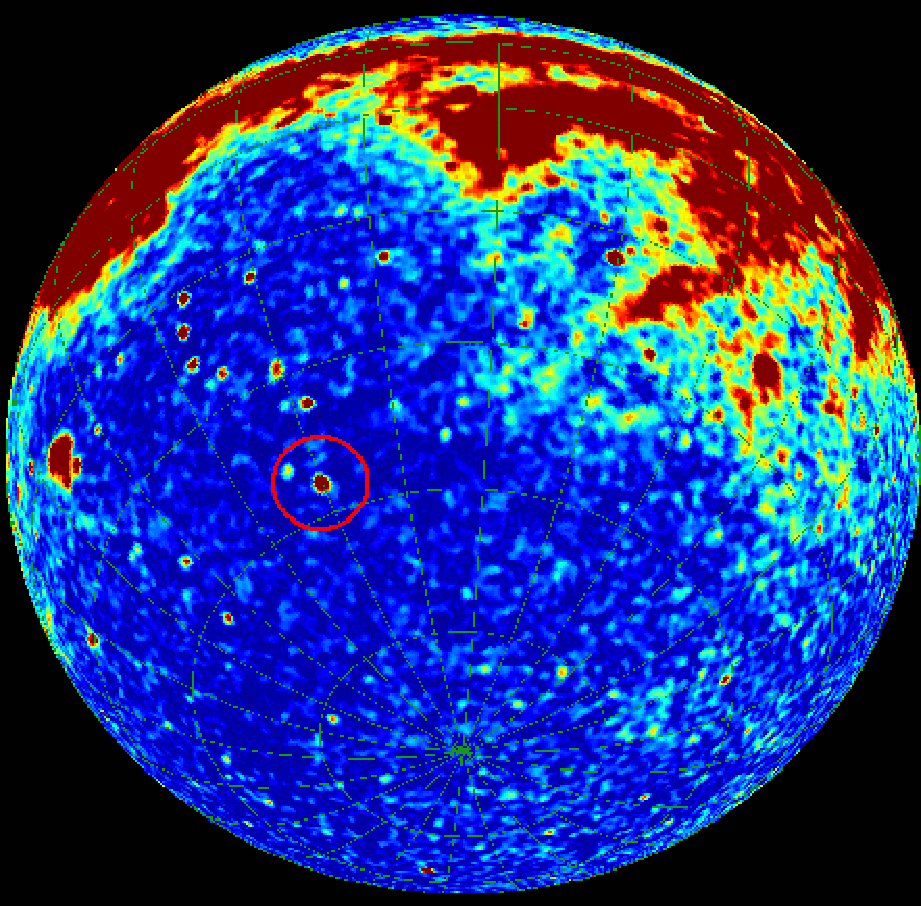}
\epsfxsize=0.45\columnwidth \epsfbox{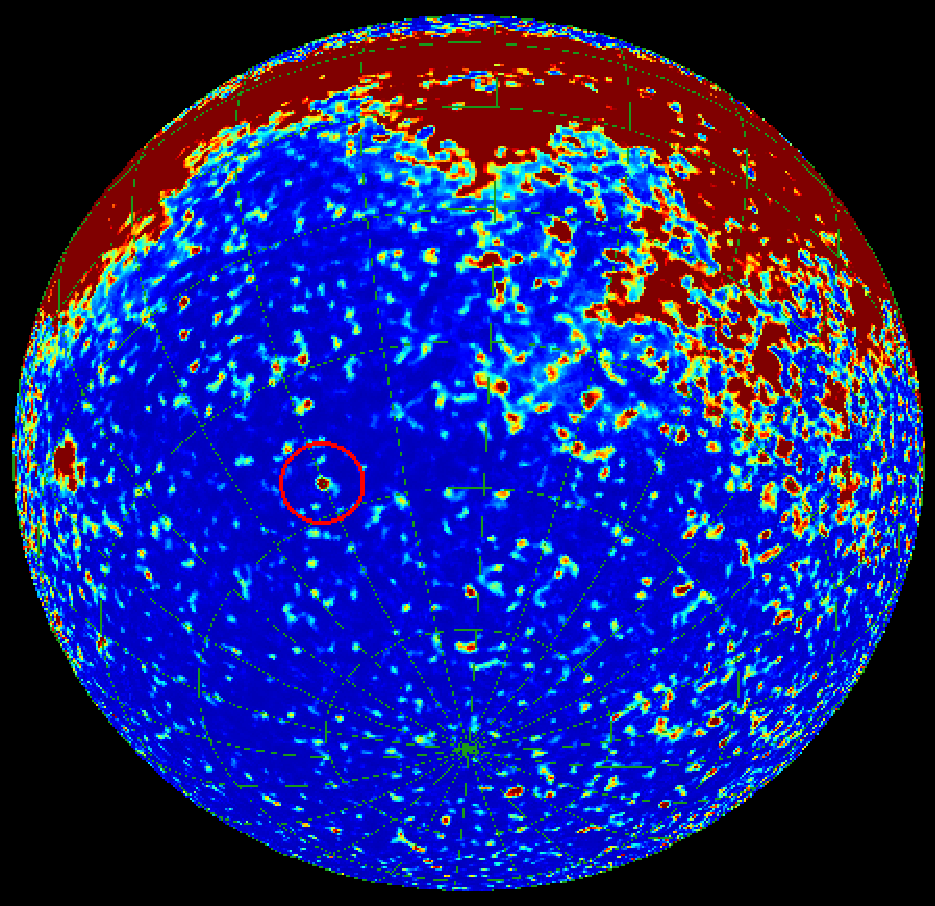}
\caption{The Ka (left) and W (right) band signal from the South hemisphere.
The red circle marks the location of the point source at the edge of the CS.
}
\label{dddem}
\end{figure}

However, from the middle plot of Fig.\ref{dem} we can see that  the
CS, as well as all the large clusters come from the smoothed signal,
which preserves all the  non-Gaussian features mentioned in the
previous section including the CS. The bottom plot of Fig.\ref{dem}
shows one more significant peculiarity of the CS zone. Just at the
edge of the CS one can see a bright point source. The amplitude of
this source is about $50\sigma_{cmb}$ for the K band and it drops
down to $6\sigma_{cmb}$ for V and W bands. In Fig.\ref{dddem} we
show this point source for the Ka and W bands. For the ILC III and
WCM maps the residuals from this point source are associated with
local maximum at the level of $1\sigma_{\rm cmb}$.

\section{Are there other ``cold'' or ``hot'' spots ?}

\begin{figure}
\setcaptionmargin{5mm}
\onelinecaptionstrue
\centering
\vspace{0.01cm}\hspace{-0.1cm}\epsfxsize=16cm
\epsfxsize=0.7\columnwidth \epsfbox{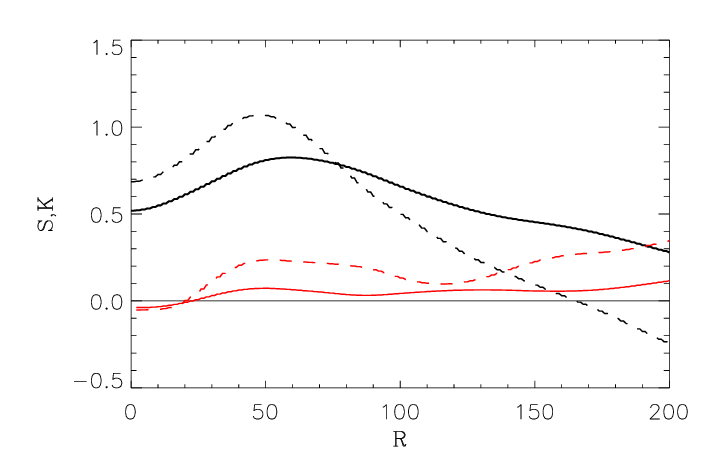}
\epsfxsize=0.7\columnwidth \epsfbox{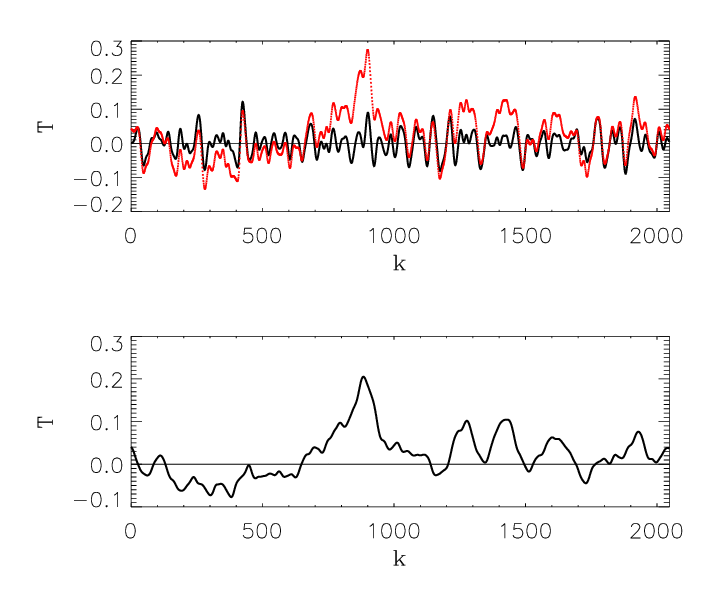}
\caption{Top: The skewness and kurtosis for the whole $b=-70^\circ$  ring.
The definition of the lines is the same as in Fig.\ref{dem}(top).
Middle: The ILC III ring $b=-70^\circ$ before demodulation
(the
grey
dots) and
after (the black line). Bottom. The ILC III signal, smoothed over
$R=50$ pixels (the corresponding angular scale $\Delta\phi=8.8^\circ$.)
}
\label{dem70}
\end{figure}

This question seems to be one of the most important question of the
entire analysis of the origin of the CS. Do we have the only one
single peculiar zone in the CMB map, or are  there more zones with
non-Gaussian properties of the signal~? To answer this question, we
suggest using the distribution of the variance of the CMB signal
versus the latitude, shown in Fig.5. Let us focus attention on the
Southern hemisphere and in particular on the point of maximum of the
variance. We can see the following coordinates of the points of
maxima: $\theta=-80^\circ,-70^\circ,-57^\circ,-30^\circ$. For these
rings we  performed the same analysis as for the ring at
$b=-57^\circ$. All of them show the existence of non-Gaussian
features, which can be identified by the same method as we have used
for the $b=-57^\circ$ ring. In Fig.\ref{dem70} we show the signal
for the ring $-70^\circ$. A non-Gaussian hot spot clearly seen at
$k\sim 850$ (the corresponding Galactic coordinates are
$b=-70^\circ, \phi=149^\circ$). The amplitude of the peak is about
$4.8\sigma_o$, and it is the member of the cluster with $d^{+}=6$
for $\nu_t=0$. From Fig.\,\ref{dem70} one can obtain the optimal
size of the filter scale $R\simeq 50$.

\begin{figure}
\setcaptionmargin{5mm}
\onelinecaptionstrue
\centering
\vspace{0.01cm}\hspace{-0.1cm}\epsfxsize=16cm
\epsfxsize=0.7\columnwidth \epsfbox{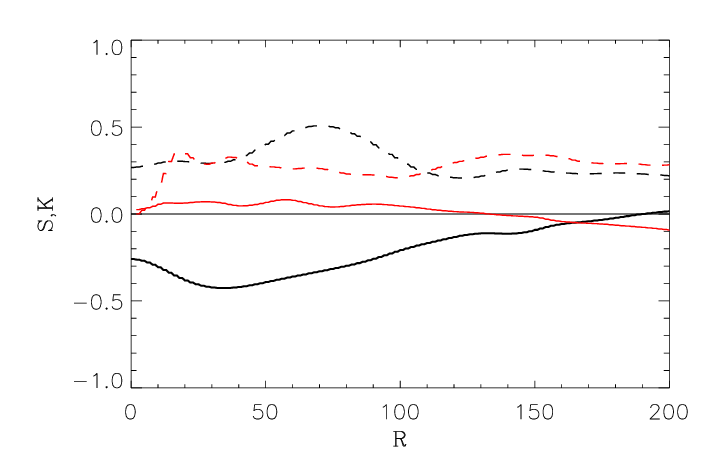}
\epsfxsize=0.7\columnwidth \epsfbox{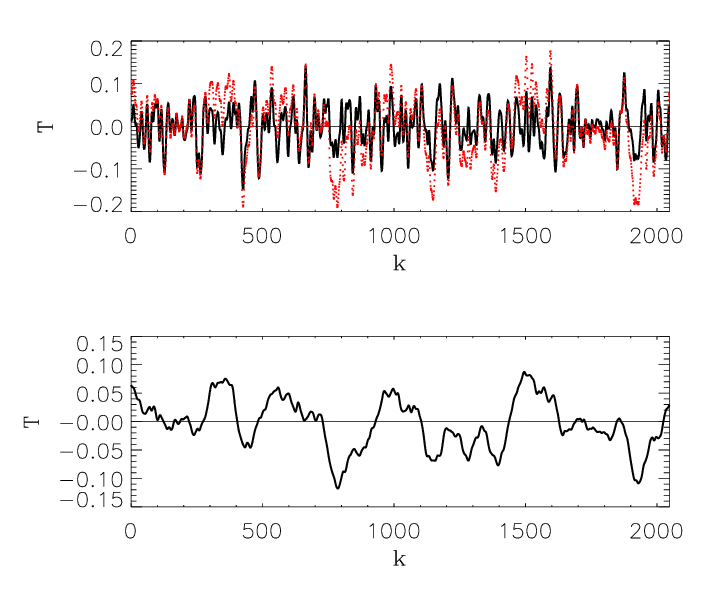}
\caption{The same as Fig.\ref{dem}, but for $b=-30^\circ$ ring.
The scale of filtration is $R=70$ pixels.
}
\label{dem40}
\end{figure}

\begin{figure}
\setcaptionmargin{5mm}
\onelinecaptionstrue
\centering
\vspace{0.01cm}\hspace{-0.1cm}\epsfxsize=16cm
\epsfxsize=0.7\columnwidth \epsfbox{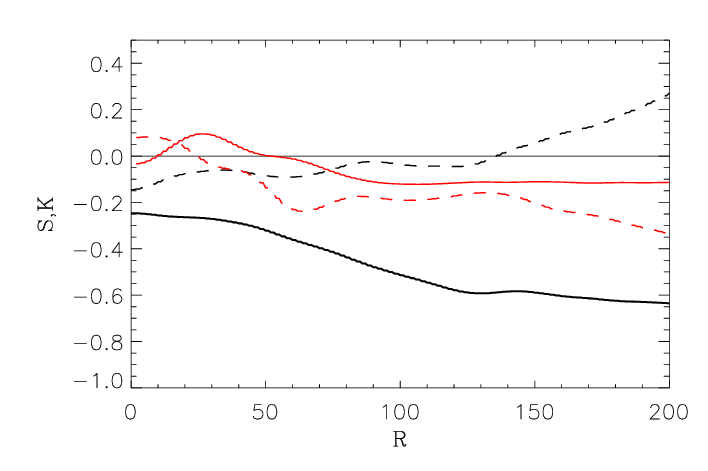}
\epsfxsize=0.7\columnwidth \epsfbox{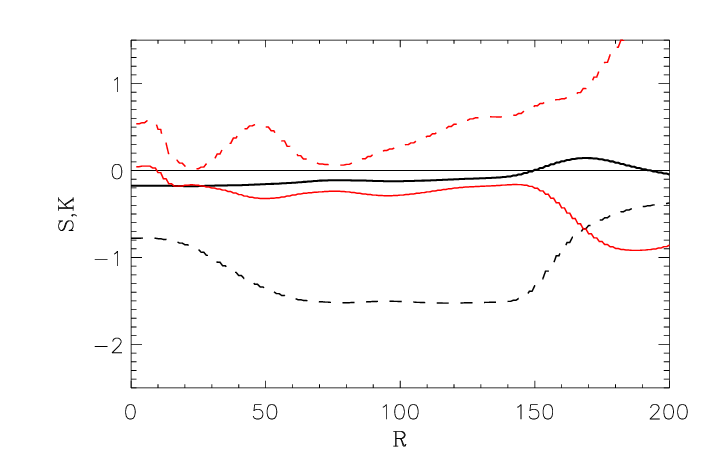}
\caption{Skewness and kurtosis for the $b=57^\circ$ ring. Top plot show
these characteristics for the whole ring, bottom plot is for
the combination of the clusters $C^{-}15+C^{+}16$.
The color of the lines is the same as for Fig.\,9.
}
\label{dem57}
\end{figure}

 Fig.\ref{dem40} shows the same characteristics of the signal as the previous
one, but for the $b=-30^\circ$ ring. Once again, the characteristic
scale of filter is about $R=70$ pixels. One may continue the search
for other rings belonging to the Southern hemisphere, just by
following the distribution of maxima of the variance from Fig.5.
However, the question is could we find the same peculiarities of the
signal for the Northern hemisphere, where a deficit of the variance
occurs? In Section 3 we already mentioned that the ring $b=57^\circ$
does indeed reveal over-clustering. Let us look closely at the
skewness and kurtosis for that ring and find out the characteristic
scale of non-Gaussianity. What is interesting is that for the
$b=57^\circ$ ring the skewness and kurtosis shown in
Fig.\,\ref{dem57} are very close to the Gaussian characteristics,
while for the zone $k_{\rm min}=1520,k_{\rm max}=1940$ with clusters
$C^{-}15,k_{\rm min}=1520, k_{\rm max}=1731$ and $C^{+}16,k_{\rm
min}=1732, k_{\rm max}=1940$ we can clearly see two minima for the
demodulated signal, one at $R=20$ and another at $R=70$.

However, we stress that for the whole ring at $b=57^\circ$ even
without filtering, the skewness and kurtosis are close to the GRF,
unlike, for example $b=-57^\circ$ ring. Moreover, increasing the
scale $R$ results in greater departures from Gaussianity, as it seen
in Fig.\,\ref{dem57}. This tendency is common for $b=72^\circ$ ring
(the point of local minimum of the variance), as for the $b=78^\circ$
(the point of local maximum of the variance from Fig.\,5). Thus,
significant non-Gaussianity of the ILC III and WCM maps mainly
corresponds to the South hemisphere, and is associated with large
angular scales, around $9-12^\circ$.

\begin{figure}
\setcaptionmargin{5mm}
\onelinecaptionstrue
\centering
\vspace{0.04cm}\hspace{-0.1cm}\epsfxsize=16cm
\epsfxsize=0.5\columnwidth \epsfbox{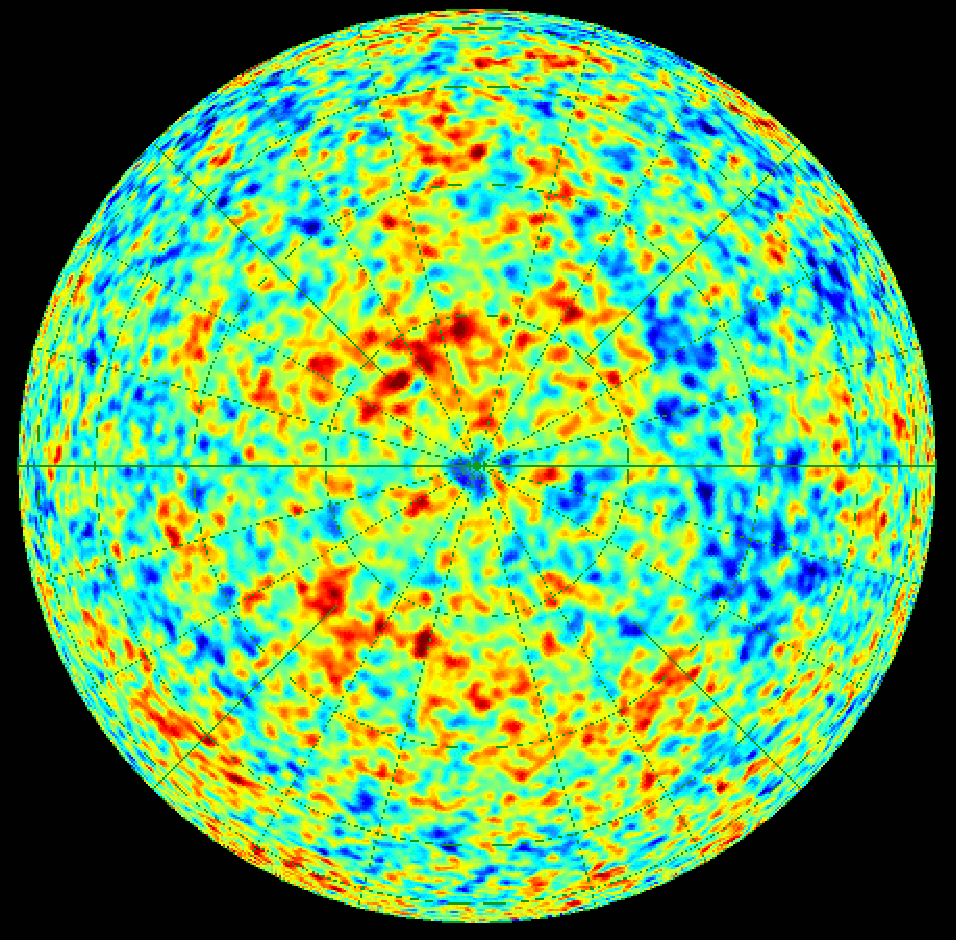}\hfill
\epsfxsize=0.5\columnwidth \epsfbox{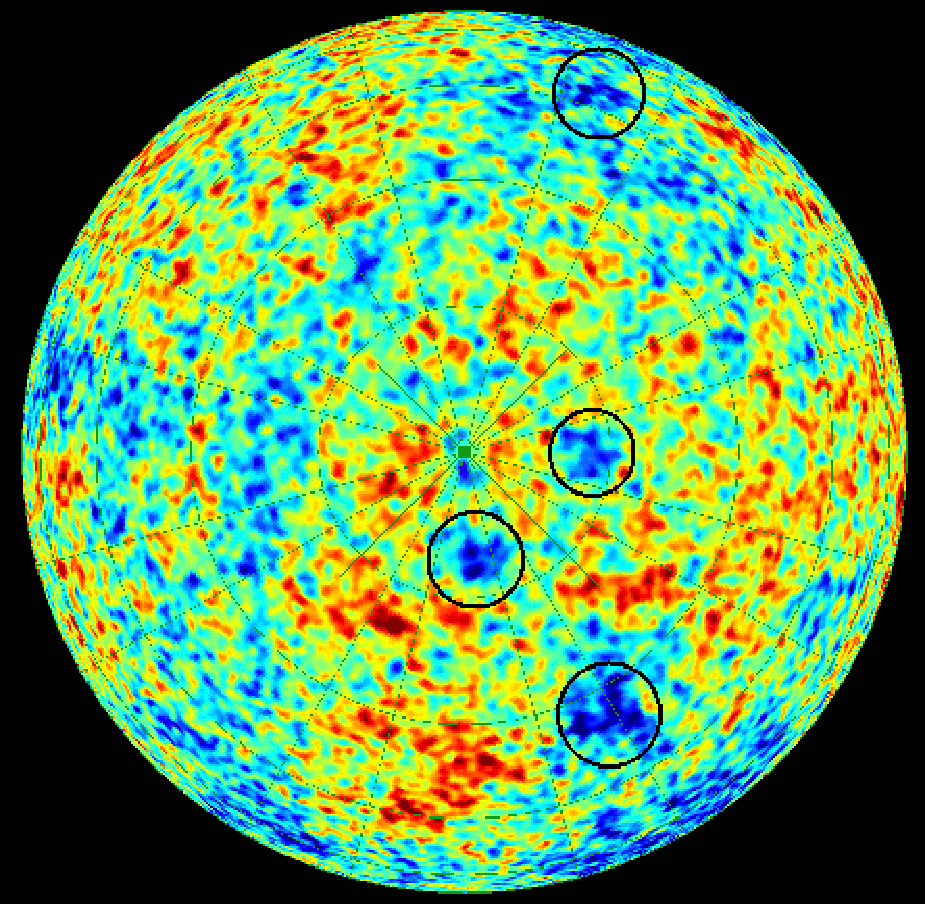}
\centering
\vspace{0.04cm}\hspace{-0.1cm}\epsfxsize=16cm
\epsfxsize=0.5\columnwidth \epsfbox{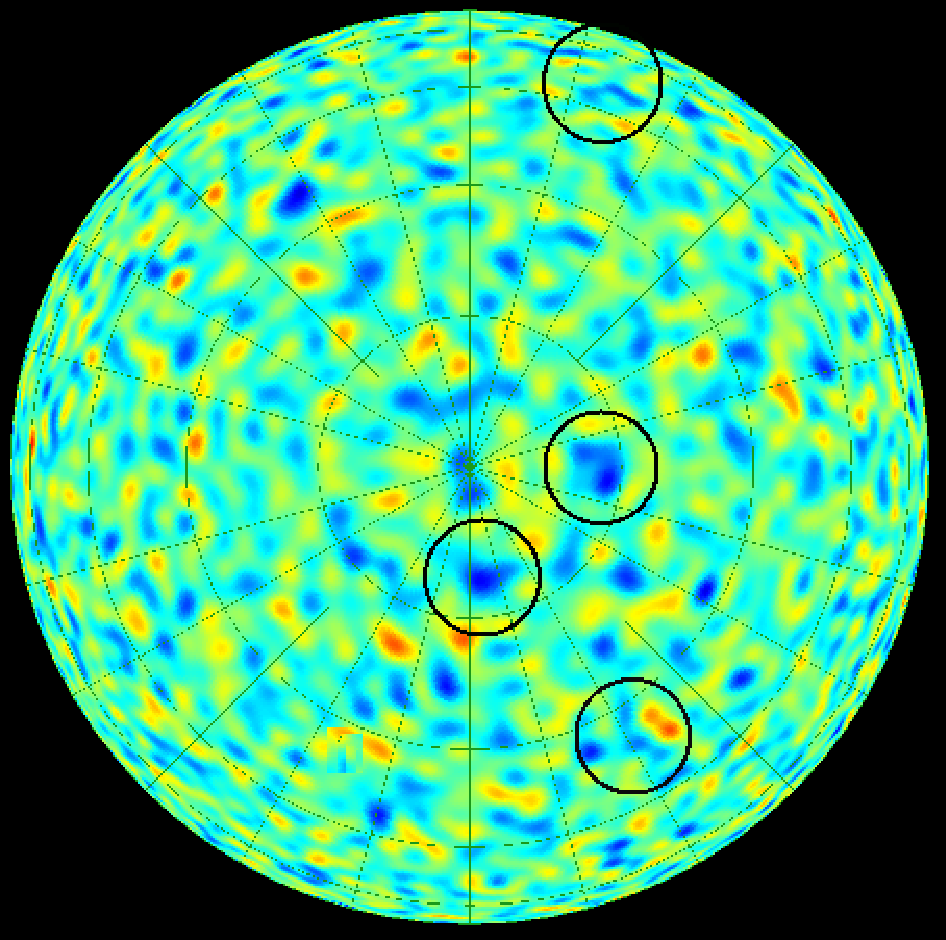}
\caption{The image of the sky seen from the North and the South galactic poles
(top and middle panels). The black circle  marks the position of the CS.
Bottom panel shows the  South hemisphere of the CMB sky after subtraction
of the first 20 multipoles of the ILC III signal.
}
\label{fsouth}
\end{figure}

To show that clustering of the extrema of the ILC III and WCM
signals is a typical feature of the morphology, we show in
Fig.\ref{fsouth} the ILC III  map seen from the North and the
South Galactic poles. From the ring with  the latitude $b\simeq
80^\circ$ one can easily find a big cluster of maxima  and one big
cluster nearly at the same latitude. For the South pole the zone of
the CS connects with three big clusters of maxima.

\begin{figure}
\setcaptionmargin{5mm}
\onelinecaptionstrue
\centering
\vspace{0.01cm}\hspace{-0.1cm}\epsfxsize=16cm
\epsfxsize=0.55\columnwidth \epsfbox{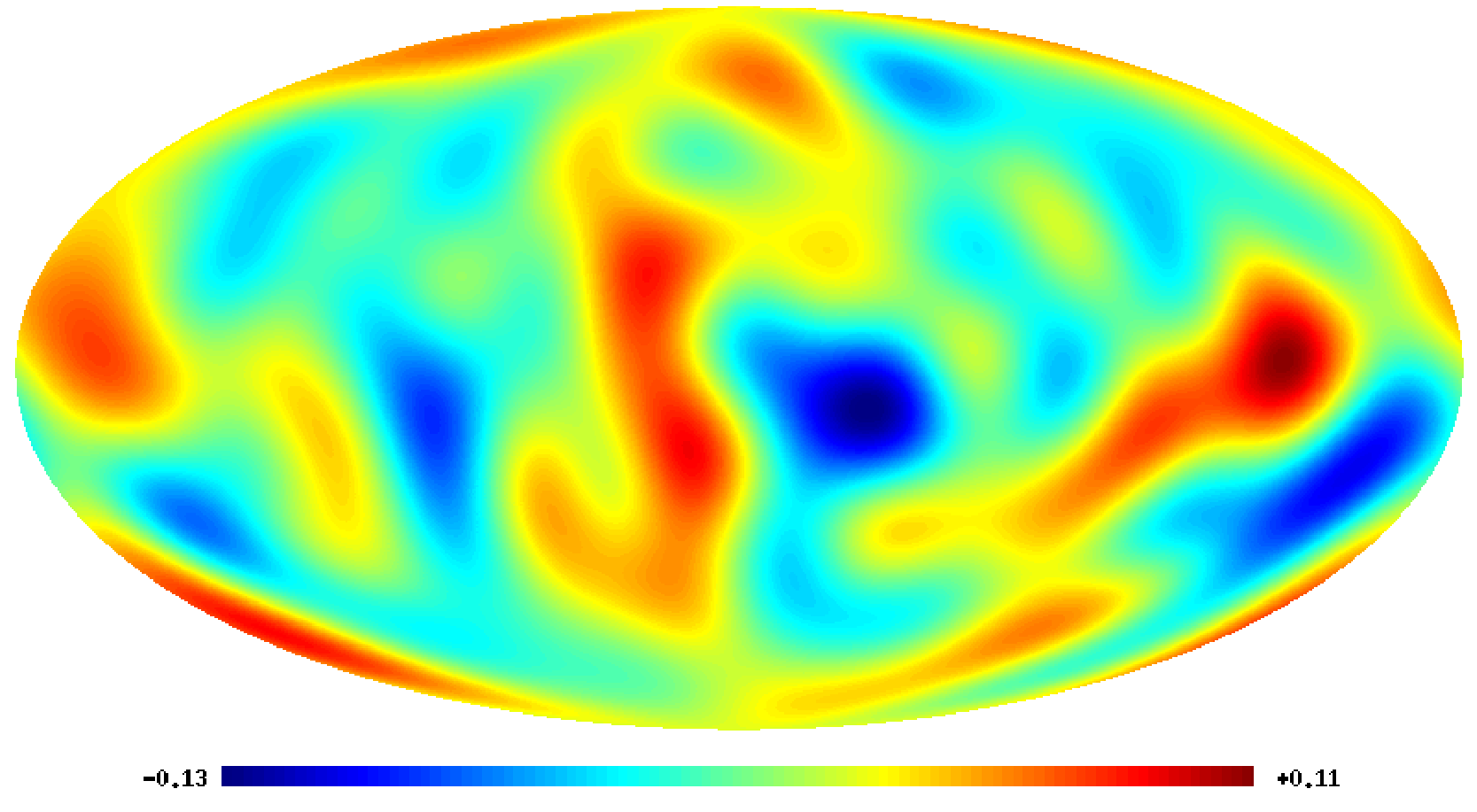}
\vspace{0.01cm}\hspace{-0.1cm}\epsfxsize=16cm
\epsfxsize=0.55\columnwidth \epsfbox{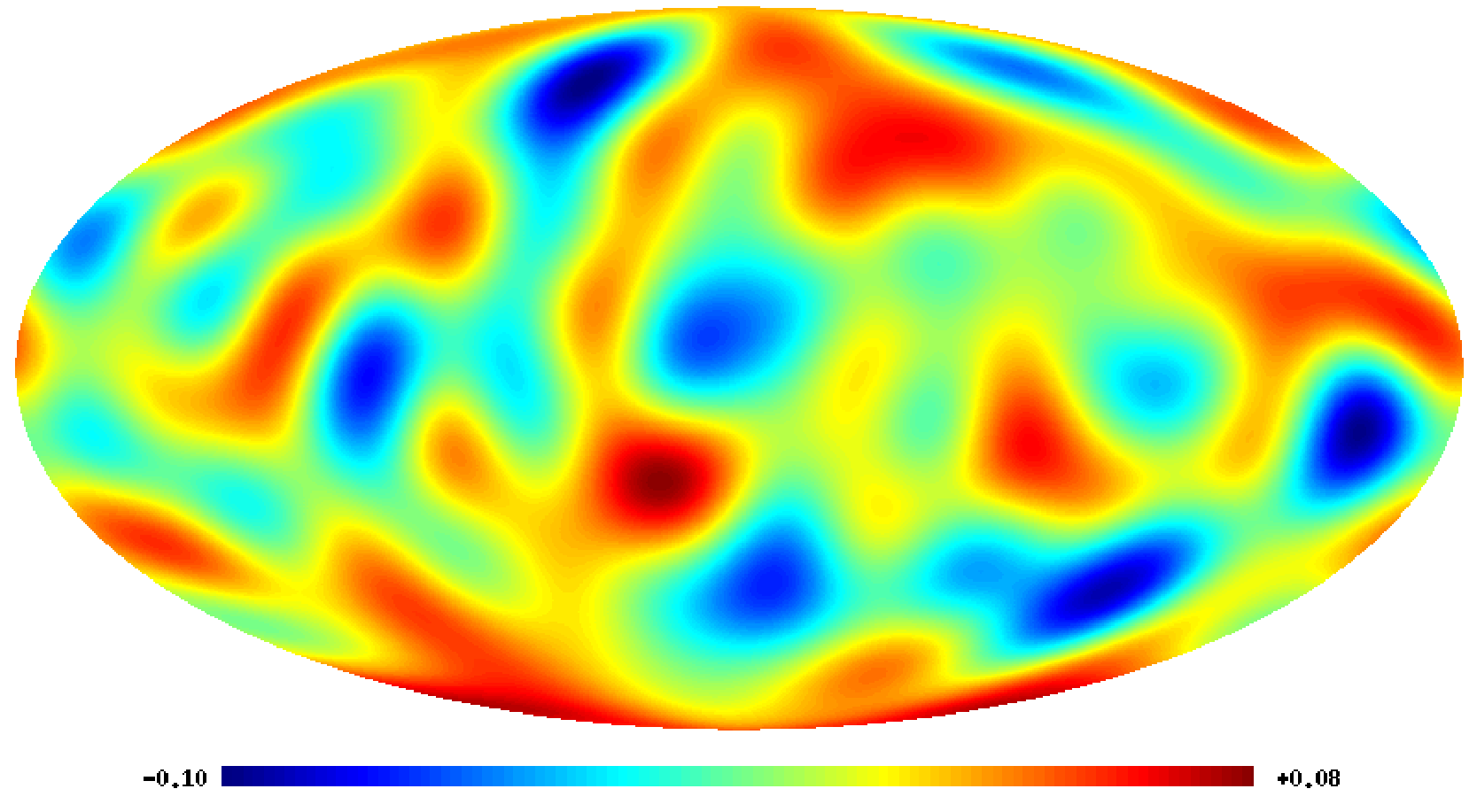}
\vspace{0.01cm}\hspace{-0.1cm}\epsfxsize=16cm
\epsfxsize=0.55\columnwidth \epsfbox{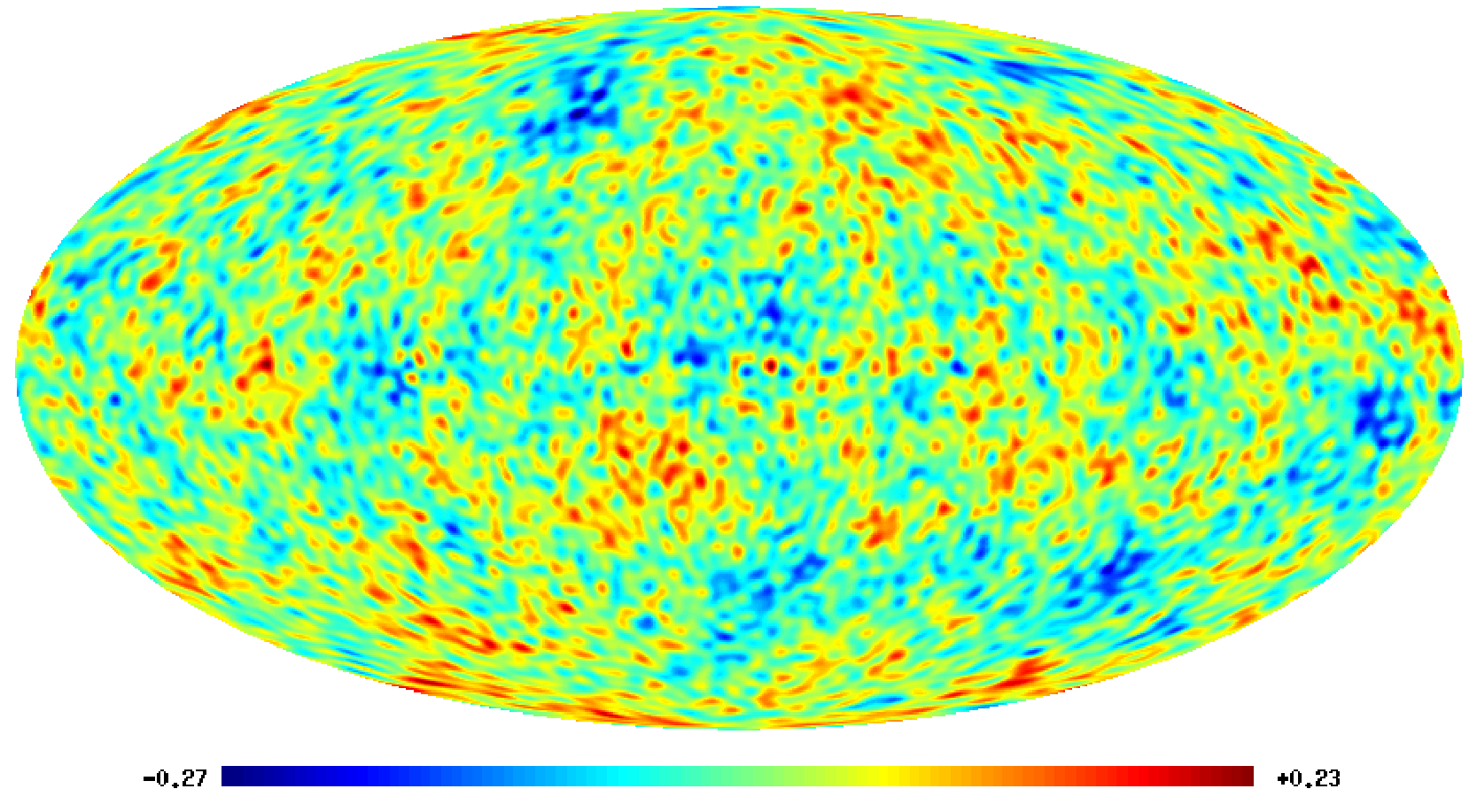}
\vspace{0.01cm}\hspace{-0.1cm}\epsfxsize=16cm
\epsfxsize=0.55\columnwidth \epsfbox{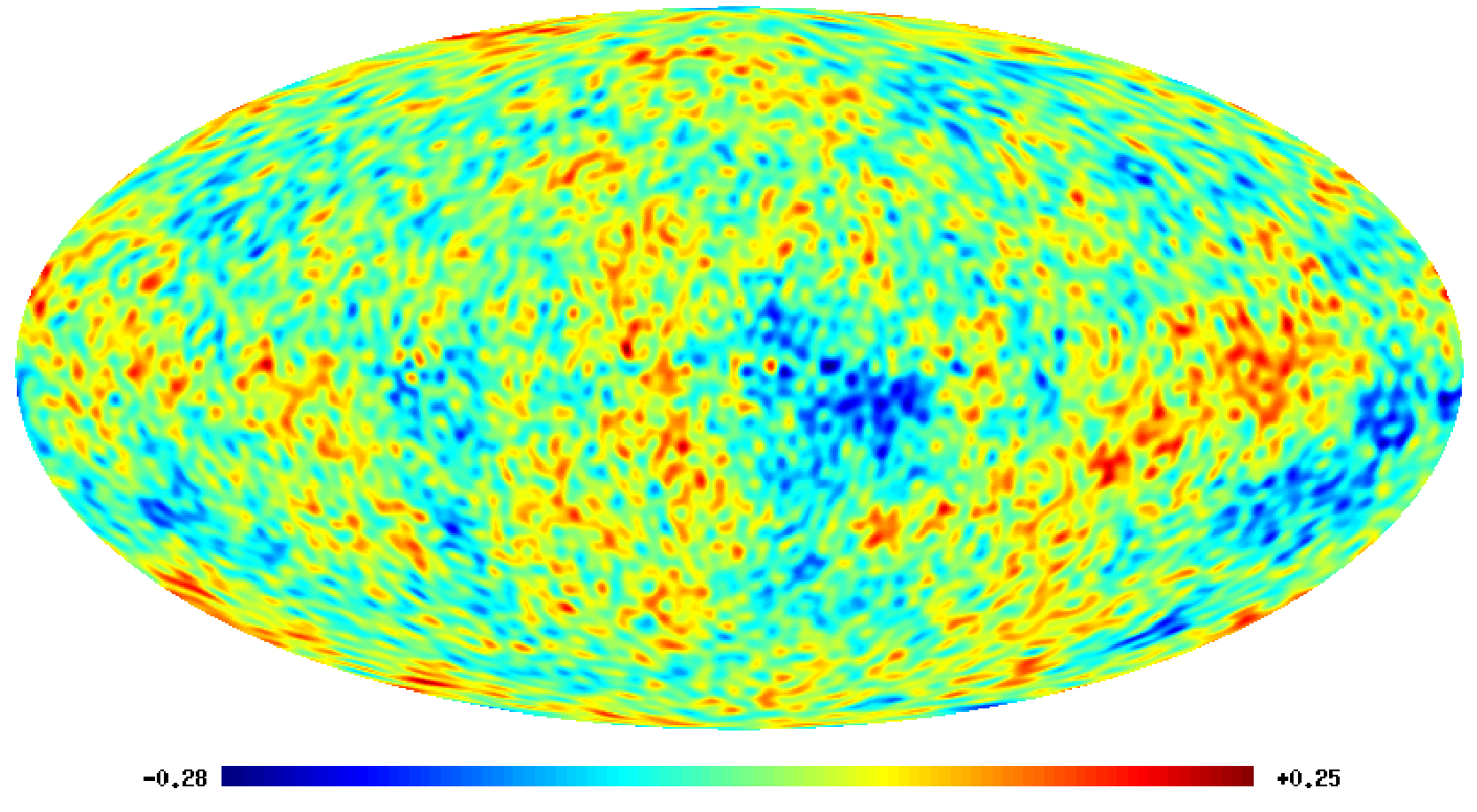}
\caption{Top. The ILC III map for $\l\le 10$. The second one from the
top is the map for RGF CMB with the same resolution as the top map.
The second one from the bottom panel shows the combined ILC signal with first
10 multipoles taking from the random signal map (middle panel), and
the rest of the multipoles $11\le \l \le 100$ from the ILC III map.
The bottom map is the ILC III with $2\le \l \le 100$.
}
\label{comb}
\end{figure}

At the end of this section we would like to give one more
argument that the over-clustering of the WCM and the ILC III maps
can be related to the peculiar properties of the WMAP low angular
resolution signal. In particular, we will show, that by changing
the properties of the first 10 harmonic of the signal, we can easily
destroy the non-Gaussianity of the CS. To show that we took under
consideration the $a_\lm$ coefficients of the ILC III map for $\l\le
10$ and replace them by the $a^{rand}_\lm$ coefficients, taking from
the random generator. The rest of the $a_\lm$ for $11\le \l \le 100$
corresponds to the ILC\,III signal.  If the non-Gaussianity of the CS
is related to high multipoles, this replacement should not change
significantly the statistical properties of the signal in the zone
around the CS. However, if the non-Gaussianity of the CS is related
to the correlations between low multipoles ($\l\le 10$), the
peculiarity of the CS should be broken. In Fig.\ref{comb} we plot
the ILC III map for $\l \le 10$, the RGF CMB map with the same
resolution and the combined map in which the first 10 the ILC III
coefficients $a_\lm$ were replaced by the coefficients
$a^{rand}_\lm$.

One can see, that the CS zone is still visible in the combined map,
but the amplitude and the morphology of the CS are changed
dramatically. Firstly, the CS is no longer one of the deepest minima
in the map. The amplitude of the CS drops down to $-0.15$\,mK from
$-0.22$\,mK. Secondly, the morphology of the zone around the CS is no
longer corresponds to the cluster with dimension 2, surrounded by
the clusters with dimension 5. The morphology of this zone now is
close to the morphology of the signal shown in Fig.10 (top panel,
the black solid line). This result is not surprising at all. As it
was shown by
\cite{bond_efsta}  
the CMB signal can be
represented as a set of peaks with different amplitudes. For the RGF
CMB signal the number density of these peaks and the shape of each
peak in the vicinity of maxima depends on spectral parameters
$\sigma_0,\sigma_1$ and $\sigma_2$, where

\begin{eqnarray}
 \sigma^2_0\sim\frac{1}{4\pi}\sum_m(2\l+1)C(l),
\hspace{0.5cm}
\sigma^2_1\sim\frac{1}{4\pi}\sum_m(2\l+1)\l(\l+1)C(l),\nonumber\\
\sigma^2_2\sim\frac{1}{4\pi}\sum_m(2\l+1)\l^4C(l),
\hspace{0.5cm}R_1=\frac{\sigma_1}{\sigma_2},
\hspace{0.5cm}\gamma=\frac{\sigma^2_1}{\sigma_0\sigma_2},
\hspace{0.5cm}\nonumber\\
\label{peaks}
\end{eqnarray}
 where $C(\l)$ is the power spectrum of the RGF CMB. It is well known that the
 parameter $R_1$ determines the correlation length of the CMB signal,
 which is proportional to the length of the cluster with dimension 1
 ($\nu_t=0,\nu=1$).  As it is seen from Eq.(\ref{peaks},
for $C(\l)\propto1/\l(\l+1)$,
 this parameter mainly depends on the maximal resolution of the map.
However, the
 clustering of the peaks is determined by the parameter $\gamma$
\cite{novjorg}  
 This parameter depends on the variance $\sigma_0$, which
 for $C(\l)\propto1/\l(\l+1)$ has logarithmic behavior.
Thus, the value of $\sigma_0$
 depends on the low multipole part of the power spectrum and consequently,
 the $\gamma$-parameter reflects directly this tendency. Thus,
 replacing of the low multipole part of the ILC III signal
by $a^{rand}_{\lm}$ from the random Gaussian signal
breaks down the asymmetry of the power spectrum of the combined ILC
map.

Moreover, in the combined high resolution map the deepest negative
peaks are related to the negative peaks, clearly seen in the RGF CMB
low resolution map (the second from the top panel). This simplest
method confirms our conclusion that peculiar structure of the
cluster around the CS is determined by the low multipole part of the
ILC III signal, which reveals significant peculiarities.

\section{Comparison with  peculiar zones detected by wavelets}

Detection of peculiarities in the CMB sky is obviously one of the
major steps in the investigation of the departure of the signal from
statistical isotropy and homogeneity. As was mentioned in the
introduction, this problem was discussed
in
\cite{mcewen2006b}
where Spherical Mexican Hat Wavelet (SMWH), elliptical SMHW, and
Spherical Butterfly Wavelet (SBW) approaches were applied to the
detection and location of the position of the  spots in
the CMB sky correlated with NVSS point sources catalog.
Below we refer to these methods collectively as wavelet
methods.
McEwen et al. 
\cite{mcewen2006b}
summarize the detection of peculiar
zones with approximate (estimated) coordinates of these zones. We
present these data in Table\,1 in order to compare these results
with our results of detection of the high dimensional clusters .

\begin{table}
\label{temperature}
\begin{center}
\setcaptionmargin{5mm}
\onelinecaptionstrue
\caption{The peculiar zones of the CMB sky detected by wavelets
(first and second columns) from
McEwen et al. (2006b).
The second
column shows the longitude $\phi$ and the latitude $\theta$ of each
zone. The third column shows the longitudes of the clusters where
the wavelet peculiar point is the member. The forth column shows the
number of extrema in the cluster ($S$) and the length of the cluster
($D$). The sign $``-''$ means that the wavelet peculiar zone is not
detected by CA with $\nu_t=0$. Stars redirect the reader to
Fig.\,\ref{fcf}.
 }
{\tiny
\begin{tabular}{|c|r|c|c|}
\hline
 Zone & Location, W & Location, CA &$S$, $D$ \\
  & $\phi$, $\theta$ (deg)&  $\phi_{min}$, $\phi_{max}$& \\
\hline
\hline
1 & 75, 57  &  56, 86  & 3, 0.083 \\
  &         & 157, 183 & 3, 0.072 \\
  &         & 267, 304 & 4, 0.102 \\
  &         & 304, 341 & 4, 0.102 \\
\hline
2 & 75, 53  &  69,  93 & 2, 0.067 \\
  &         &  40,  66 & 4, 0.072 \\
  &         & 160, 183 & 3, 0.064 \\
  &         & 273, 310 & 4, 0.103 \\
\hline

3 & 323, 56 & 304, 339 & 4, 0.098 \\
  &         & 278, 304 & 3, 0.072 \\
\hline
4 & 321, 62 & 306, 333 & 2, 0.075 \\
  &         & 333, 357 & 3, 0.067 \\
  &         & 272, 307 & 4, 0.097 \\
\hline
5 & 267, 50 & 251, 274 & 3, 0.064 \\
  &         &  58,  82 & 3, 0.067 \\
  &         & 313, 344 & 4, 0.086 \\
\hline
6 & 268, 45 & $-$      & $-$      \\
  &         & 129, 156 & 3, 0.075 \\
\hline
7 & 213, 40 & $-$      & $-$      \\
  &         &  30,  58 & 4, 0.078 \\
  &         & 233, 258 & 4, 0.069 \\
\hline
8 & 223, 30 & $-$      & $-$      \\
  &         &   9,  37 & 4, 0.078 \\
  &         & 136, 157 & 4, 0.056 \\
\hline
9 & 160, 26 & $-$      & $-$      \\
  &         & 174, 213 & 6, 0.108 \\
  &         &   4,  32 & 4, 0.078 \\
\hline
10&  94, -28 & $-$      & $-$       \\
  &          &  35,  62 &  3, 0.075 \\
  &          & 142, 235 & 15, 0.258 \\
  &          & 333, 352 &  4, 0.056 \\
\hline
11&  81, -34 & $-$      & $-$        \\
  &          & 141, 181 &  7, 0.11  \\
  &          & 187, 230 &  6, 0.12  \\
  &          & 261, 294 &  7, 0.092 \\
\hline
12& 118, -42 & -         & -        \\
  &          &   8,  30  & 4, 0.061 \\
  &          & 333,360+8 & 5, 0.097 \\
\hline
13&  20, -48 &   7,  30  & 4, 0.058 \\
  &          &  70, 105  & 4, 0.097 \\
  &          & 243, 278  & 4, 0.097 \\
\hline
14&  34, -31 &  32,  55  & 5, 0.064 \\
  &          & 288, 330  & 6, 0.117 \\
\hline
15& 230, -68 & 220, 253  & 2, 0.092 \\
  &          &  19,  52  & 2, 0.092 \\
  &          &  63, 117  & 5, 0.15  \\
\hline
16& 204, -56 & 198, 216  & 2, 0.05$^*$ \\
  &          & 148, 188  & 4, 0.11  \\
  &          & 223, 270  & 5, 0.13  \\
\hline
17& 186, -54 & 193, 216  & 2, 0.063$^*$\\
  &          & 150, 193  & 4, 0.12  \\
  &          & 224, 294  & 6, 0.194 \\
\hline
18& 218, -33 & 187, 230  & 3, 0.12   \\
  &          & 258, 294  & 6, 0.10$^*$ \\
\hline
\end{tabular}
}
\end{center}
\end{table}

\begin{figure}
\setcaptionmargin{5mm}
\onelinecaptionstrue
\centering
\vspace{0.01cm}\hspace{-0.1cm}\epsfxsize=16cm
\epsfxsize=0.4\columnwidth \epsfbox{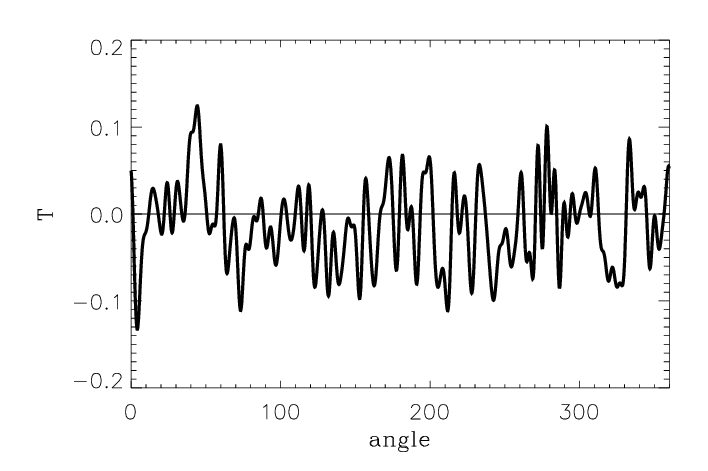}
\epsfxsize=0.4\columnwidth \epsfbox{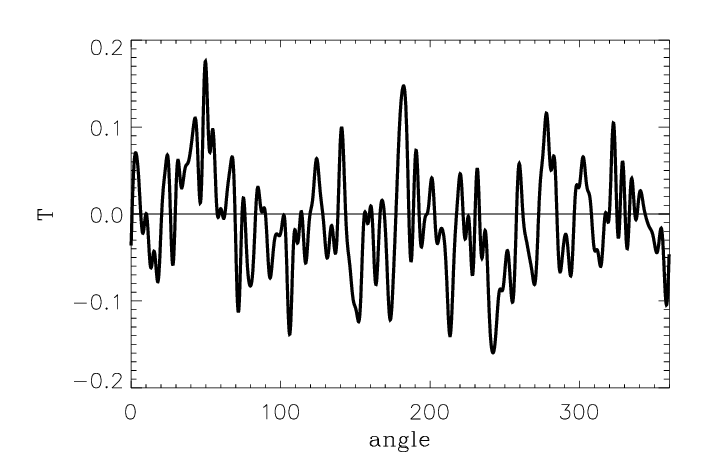}
\centering
\vspace{0.01cm}\hspace{-0.1cm}\epsfxsize=16cm
\epsfxsize=0.4\columnwidth \epsfbox{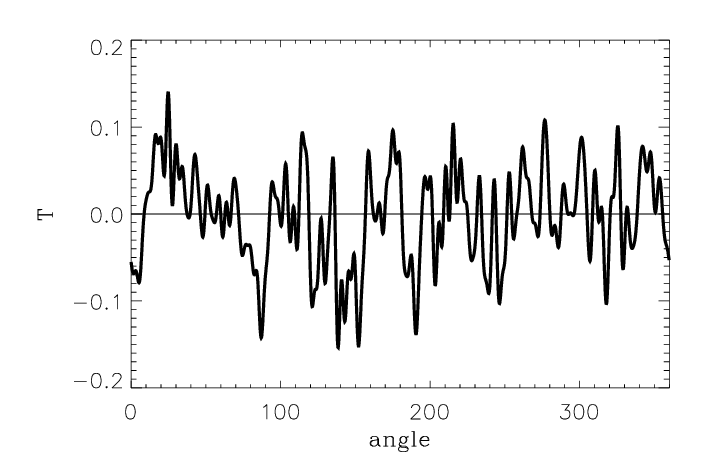}
\epsfxsize=0.4\columnwidth \epsfbox{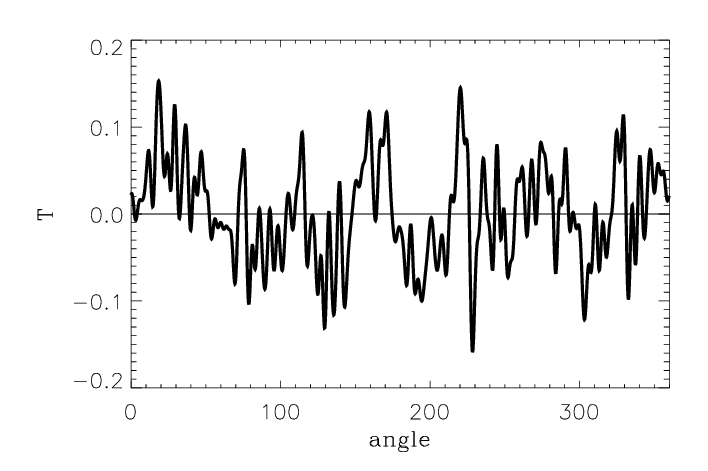}
\centering
\vspace{0.01cm}\hspace{-0.1cm}\epsfxsize=16cm
\epsfxsize=0.4\columnwidth \epsfbox{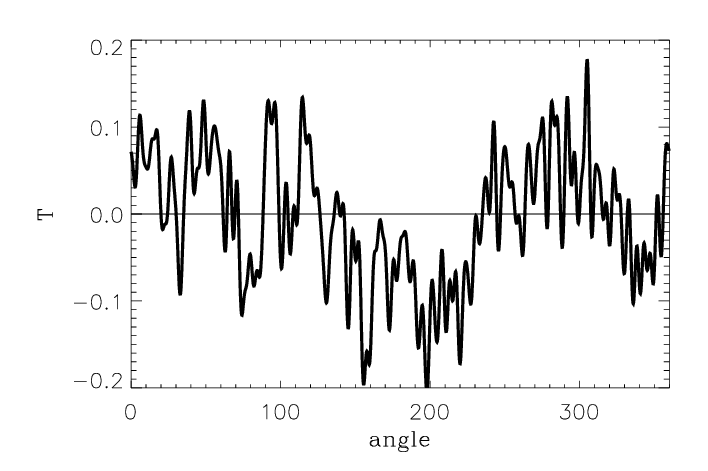}
\epsfxsize=0.4\columnwidth \epsfbox{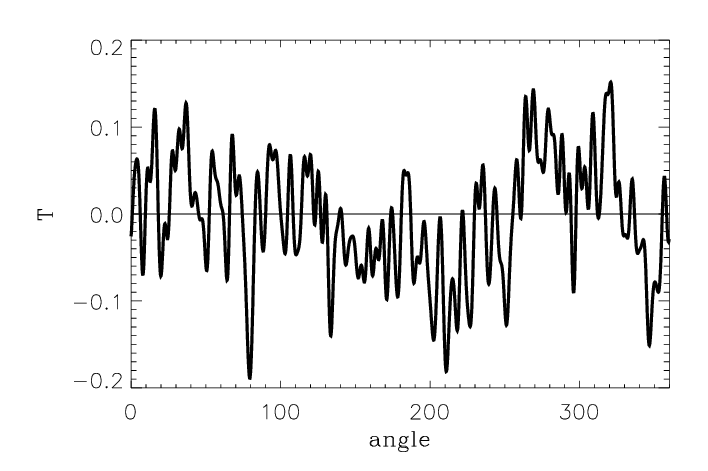}
\centering
\vspace{0.01cm}\hspace{-0.1cm}\epsfxsize=16cm
\epsfxsize=0.4\columnwidth \epsfbox{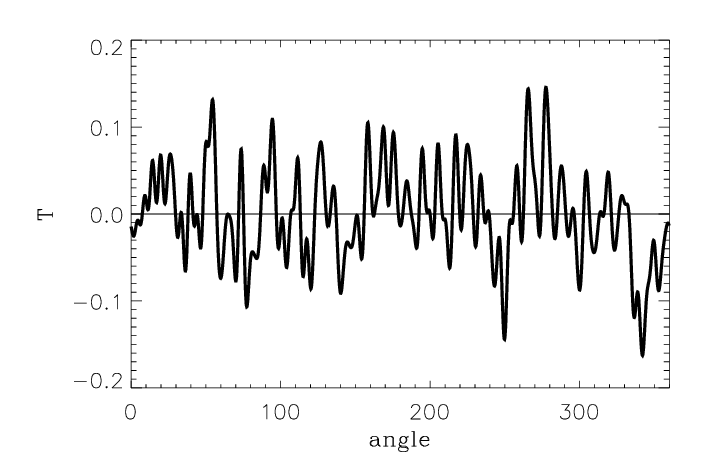}
\epsfxsize=0.4\columnwidth \epsfbox{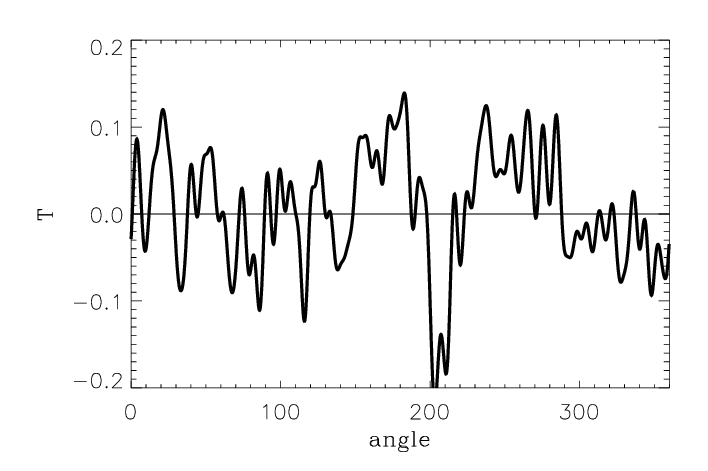}
\centering
\vspace{0.01cm}\hspace{-0.1cm}\epsfxsize=16cm
\epsfxsize=0.4\columnwidth \epsfbox{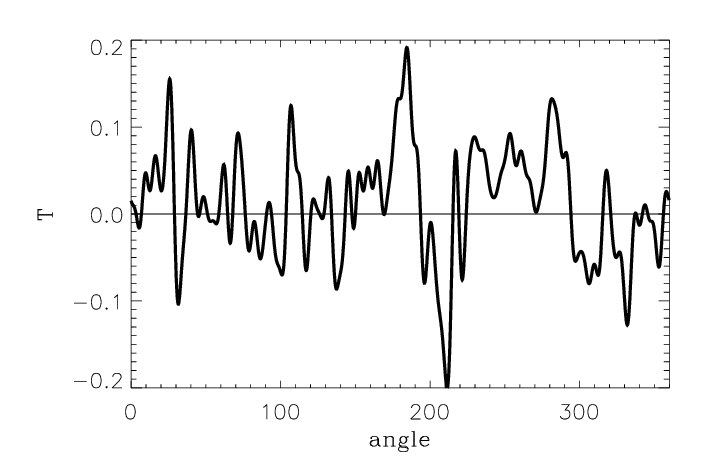}
\epsfxsize=0.4\columnwidth \epsfbox{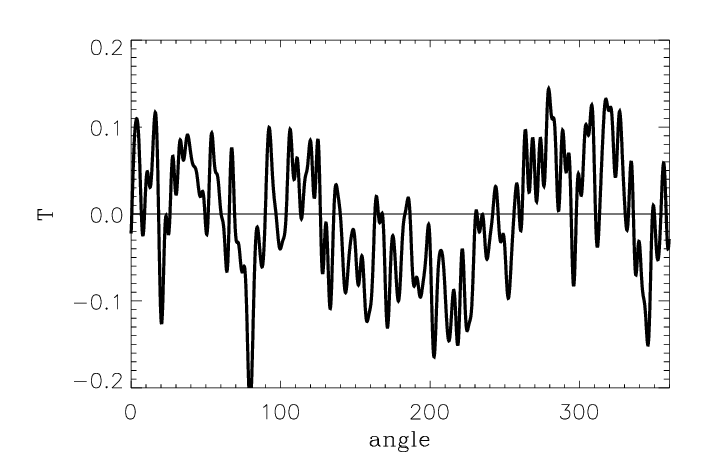}
\caption{The CMB signal for equal latitude rings, marked in Table\,1
by sign ``$-$''. From the top to the bottom and from the left
to the right are the zones 6-12,16-18.
}
\label{fcf}
\vspace*{2ex}
\end{figure}
Let us discuss the properties of the signal for Zone 6 of Table\,1.
The wavelet approach gives us the azimuthal coordinate of the
 spot $\phi_6=268^\circ$. From the top left plot, we can
see that this point corresponds to a relatively small negative peak.
The same morphology of the peaks is clearly observed at $\phi\sim
210^\circ$, and $\phi\sim 330^\circ$. Both peaks are very similar to
the CS zone, but with smaller amplitudes. These two peaks probably
contribute to the overall non-Gaussianity of the signal more
significantly than the $\phi_6=268^\circ$ peak (see Table 1).

The next zone, which was not detected by cluster analysis at
$\nu_t=0$, is Zone 7 shown on the top right plot. According to the
wavelet analysis, the azimuthal coordinate for this zone is
$\phi_7=213^\circ$. Once again, here we have a negative peak, but
the peak at $\phi_7\simeq 240^\circ$ reveals a more significant
departure from Gaussianity, it being a member of the cluster with
number of minima $N=4$.

For Zone 8, the wavelets analysis gives us $\phi_8=223^\circ$. We
show the corresponding ring in Fig.\ref{fcf} (second from the top
left plot). The detected zones manifest themselves as points of
maxima, while the negative peaks detected by cluster analysis,
listed in Table\,1 reveal significant departure from Gaussianity.
For example, the cluster of minima at $\phi\simeq 140-160^\circ$ can
be detected not only for $\nu_t=0$ threshold, but even for
$\nu_t=-2$, as a cluster of maxima and minima. Roughly speaking, all
the pictures, shown in Fig.\ref{fcf} clearly demonstrate that
implementation of cluster analysis with different thresholds $\nu_t$
allow us to detect not only  single non-Gaussian zones, but also
clusters of these zones which have non-local structure. To
illustrate this tendency, we would like to draw attention to the
bottom right picture belonging to the Zone 18. According to
Table\,1, the non-Gaussian zone is located at $\phi_{18}=218^\circ$
being the member of the cluster and detected both by wavelet and by
cluster analysis. However, there is another zone at $\phi\simeq
80^\circ$ with a morphology similar to the morphology of the CS.
Moreover, looking at the shape of the signal at $\phi>130$ one can
see the modulation of the signal by low frequency harmonics. This
type of non-Gaussianity is an argument in favor of hypothesis that
low multipoles of the CMB signal are highly non-Gaussian.

\section{ Conclusion}

We have re-examined the properties of the Internal Linear
combination WMAP CMB map and the co-added WCM map by an analysis of
the properties of the signal in the vicinity of the CS.  These two
maps of the CMB signal display remarkably similar structures on
equal latitude rings at $|b|>30^\circ$. We have re-examined the
properties  of the CS at the galactic latitude $b=-57^\circ$ and
longitude $l=209^\circ$ and shown that it is associated with the
cluster with length $D\sim 3\langle D(n)\rangle$. In addition
 to the CS, we have also found a few more zones of the CMB
signal with almost the same morphology,  at $b=57^\circ$,
$b=-80^\circ$, $b=-30^\circ$.

From an analysis of the ILC III map we have shown that the shape of
the CS is formed primarily  by the CMB signal localized in
multipoles between $10\le \l\le 20$ (corresponding to angular scales
about $5-10^\circ$), in agreement with
results in
\cite{cruz2006,cruz2007a}.
At the same time we have demonstrated that the clustering of
the peaks in the zone around the CS depends on the low
multipole tail of the ILC III
map $2\le \l\le 10$.

Taking into account that the same  modes lead to a modulation of the
whole CMB sky, we subtracted these modes from the CMB signal. The
demodulated   CMB signal  looks like a random one without
significant over-clustering in agreement with
\cite{cruz2005}. 

We have investigated the asymmetry of the variance for iso-latitude
rings in respect to the Galactic plane. The South hemisphere has
excess variance in comparison to the North hemisphere. This is why
local defects and large clusters, including the CS and its
associated cluster, are mainly concentrated in the Southern
hemisphere.

Taking all these investigations together, we believe that the
mystery of the WMAP CS directly reflects the peculiarities of the
low-multipole tail of the CMB signal, rather than a single local
(isolated) defect or a manifestation of a globally anisotropic
model. This interpretation does not preclude the possibility of an
exotic origin of the CS and related phenomena, but it does specify
more precisely what properties such explanations must generate. A
satisfactory model of the CS must explain the entire range of its
behavior rather than only one aspect.

Our final remark is related to the definition of significance of the
CS detection by different methods, based on the assumption that
Gaussian statistics apply to the observed CMB sky.
Ever since
Eriksen et al. 
\cite{eriksenasym}
showed that the distribution of the power of the CMB across the sky
is very anisotropic at the scales about $10^\circ$, it has been
clear that Gaussian statistics are no longer a valid reference for
determining the significance of this feature. Our approach to the
large-scale angular modulation of the CMB is a possible alternative
approach to this issue.

\section*{Acknowledgments}

We acknowledge the use of the Legacy Archive for Microwave Background
Data Analysis (LAMBDA\footnote{\tt http://lambda.gsfc.nasa.gov}).
We also
acknowledge the use of \healpix package
\cite{healpix}
to produce $\alm$. The \glesp package
\cite{glesp}
was used in this work.
OVV thanks Russian Foundation for Basic Researches by the grants
No\,09-07-00159, Foundation for the National Science Support
and the ``Dynasty'' Foundation.

\def\apj{ApJ}
\def\apjl{ApJL}
\def\apjs{ApJS}
\def\mn{MNRAS}
\def\mnras{MNRAS}
\def\nature{nat}
\def\aa{A\&A}
\def\prl{Phys.\ Rev.\ Lett.}
\def\prd{Phys.\ Rev.\ D}
\def\pr{Phys.\ Rep.}
\def\ijmpd{Int. J. Mod. Phys. D}

\end{document}